\documentclass[journal=apchd5,manuscript=article]{achemso}

\usepackage{chemformula} 
\usepackage[T1]{fontenc} 

\usepackage{graphicx}
\usepackage[caption=false]{subfig}
\usepackage{amsmath,amssymb,amsfonts}
\usepackage{verbatim}
\usepackage{xcolor}

\newcommand{\rminc}{\mathrm{inc}}
\newcommand{\rmsc}{\mathrm{sca}}

\newcommand{\rmo}{\mathrm{o}}
\newcommand{\rmi}{\mathrm{i}}
\newcommand{\rmd}{\mathrm{d}}

\author{Qiang Sun}
\affiliation[RMIT]
{Australian Research Council Centre of Excellence for Nanoscale Biophotonics, School of Science, RMIT University, Melbourne, VIC 3001, Australia}
\email{qiang.sun@rmit.edu.au}
\author{Kishan Dholakia}
\affiliation[St-Andrews]
{SUPA, School of Physics and Astronomy, University of St Andrews, North Haugh, Fife KY16 9SS, United Kingdom}
\alsoaffiliation[Yonsei]
{Department of Physics, College of Science, Yonsei University, Seoul 03722, South Korea}
\author{Andrew D. Greentree}
\affiliation[RMIT]
{Australian Research Council Centre of Excellence for Nanoscale Biophotonics, School of Science, RMIT University, Melbourne, VIC 3001, Australia}

\title[]
  {Optical forces and torques on eccentric nanoscale core-shell particles\footnote{Published in \emph{ACS Photonics} (https://doi.org/10.1021/acsphotonics.0c01825).}}


\begin{document}







\begin{abstract}
  The optical trapping and manipulation of small particles is an important tool for probing fluid properties at the microscale.  In particular, microrheology exploits the manipulation and rotation of micron-scale particles to probe local viscosity, especially where these properties may be perturbed as a function of their local environment, for example in the vicinity of cells.  To this end, birefringent particles are useful as they can be readily controlled using optically induced forces and torques, and thereby used to probe their local environment.  However the magnitude of optical torques that can be induced in birefringent particles is small, and a function of the particle diameter, meaning that rotational flow cannot readily be probed on length scales much small than the micron level.  Here we show modelling that demonstrates that eccentric spherical core-shell nanoparticles can be used to generate considerable optical torques.  The eccentricity is a result of the displacement of the centre of the core from the shell. Our results show that, for particles ranging from 90~nm to 180~nm in diameter, we may achieve rotation rates exceeding 800~Hz. This fills a missing size gap in the rotation of microparticles with optical forces. The diameter of particle we may rotate is almost an order of magnitude smaller than the smallest birefringent particles that have been successfully rotated to date. The rotation of eccentric core-shell nanoparticles therefore makes an important contribution to biophotonics and creates new opportunities for rheology in nanoscale environments.
\end{abstract}

\section{Introduction}

The field of optical trapping, initiated by Arthur Ashkin~\cite{Ashkin1970-ATP} received the Nobel Prize in physics in 2018. The topic area describes the control and manipulation of mesoscale particles using the momentum of light. The added feature of including light fields with spin or orbital angular momentum has fuelled the topic of rotating as well as translating trapped particles. It has been widely applied across the sciences and has made particular impacts in areas spanning fundamental physics~\cite{Ashkin1970-ABD, Spesyvtseva2016} to biological sciences~\cite{Ashkin1987, Ashkin1992-FSG, Wang1997, MacDonald2003, Pang2014, Craig2015, Ritchie2015}. This opens up the prospect of new studies including those in microfluidics, namely cellular microrheology and rotational dynamics for the burgeoning area of levitated optomechanics.

From the fundamental physics standpoint, the area of levitated optomechanics has emerged as a powerful way to explore the boundary between classical and quantum physics with mesoscopic particles well isolated from their environs~\cite{Chang2012, Gieseler2012, Monteiro2013, Neumeier2015}. In turn, this has led to the recent demonstration of cooling of a particle to the quantum ground state~\cite{Deli2020}. In other work, the motion of trapped particles has elucidated fundamental concepts for both the linear and angular momentum of light by using a trapped particle as a probe of the incident field~\cite{RodrguezSevilla2018, Stout1997, GarcesChavez2003}.

In the area of biological science, the use of optical traps as calibrated force transducers has led to the measurement of exquisite, minuscule forces associated with a  range of linear and rotary molecular motors~\cite{Arita2020}. However the challenge of understanding how cells respond to their environment requires the probing of viscosity at the nanoscale, and this length scale is not yet accessible through the transfer of sufficient optical torque to induce rotation of nanoparticles.  
Currently, birefringent particles are often used as the optical rotation probes for local rheology measurements. However, given the small difference of the refractive indices between the optical paths of birefringent particles and optical torque in proportion to $a^3 \sim a^6$ with $a$ being the characteristic size of the particle~\cite{Liu2005}, the optical torque becomes negligible as the size of the birefringent particle reduces, especially below the sub-micron scale. Smaller nano-scale probes of viscosity should unlock cellular environmental responses as they can be taken up more readily in cells, with future applications as minimally invasive probes of the intracellular environment.

From the material science point of view, there has been considerable development of nanoparticles with complex and bespoke compositions including metallic particles or  high refractive index materials. Such more complex nanoparticles offer great promise for novel applications, and the emergence of new synthesis capabilities at the mesoscale are becoming more accessible. In this domain core-shell nanoparticles~\cite{GhoshChaudhuri2011, Gawande2015, ElToni2016} can be utilised to develop novel materials to perform multiple tasks and functions.

Here we investigate the optical forces and torques acting on non-concentric asymmetrical core-shell spherical particles (eccentric spherical core-shell nanoparticles). This is the first study of its type and opens up a hitherto unrecognised area of exerting optical torques and initiating rotation for spherical particles of very small size. Although the use of spin angular momentum is already a powerful tool, standard techniques rely on a polarisation change through the trapped object, which can be prohibitively small for particles below a micron in diameter. Accordingly, rotation studies have been performed with larger particles~\cite{Arita2016}. These studies include viscosity measurements. Conversely, as we show, non-concentric core-shell spherical particle of sizes 50-500 nm in diameter may undergo rotation. In this case, the determination of viscosity is straightforward when compared to multiparticle aggregates that can be rotated but would be more complex to model~\cite{Bang2020}. This is crucial for future studies in cellular environments incompatible with larger particles, for example and benefit from a straightforward route to determine viscosity from the system.

To date, the majority of studies of the optomechanical response of trapped particles have focused on particles that comprise of a single homogenous (typically dielectric) material. By tailoring the material property of the particle new modalities may be envisaged that cannot be readily achieved using shaped light alone with dielectric objects~\cite{Spesyvtseva2016}. Core shell particles offer new opportunities for optomechanical forces. A key example is combining the advantage of core-shell particle and optical trapping, which can open up a promising direction of nanotechnology for scientific, engineering, biological and medical applications. For instance, Jannasch \emph{et al}~\cite{Jannasch2012} enhanced optical forces to the nanonewton level by coating a titania particle with a silica shell, which is useful for biological studies ~\cite{Craig2015}. Spadaro \emph{et al}~\cite{Spadaro2015} studied how the relative thickness between the core and shell can affect the optical force on a Au–PEG core–shell particle. Ali \emph{et al}~\cite{Ali2020} investigated the effects of the chirality of the core-shell particle on the optical torque under a circularly polarised beam. So far, most studies of optical trapping of concentric core-shell nanoparticles have been directed to increasing trapping force by adding a shell with different refractive index to the core particle (for example creating an anti-reflection coating). Our work is thus distinguished in this regard as we show, breaking centro-symmetry in such systems leads to increased optical torques.

In the context of optomechanics we may ask: what is the effect of breaking the restriction of centro-symmetry? A number of research studies have demonstrated the synthesis of eccentric spherical core-shell particles with different materials. For example, using gold cores, researchers have built nanoscale eccentric core-shell particles with different shells, including Au@SiO$_2$~\cite{Chen2010}, Au@TiO$_2$~\cite{Seh2011} and Au@polymer~\cite{Chen2008}. Other than using Au as the core, with a polymer as the shell, different types of semiconductor, metal and dielectric materials have also been used as the core material for the synthesis of nanometric eccentric core-shell particles, such as Fe$_2$O$_3$@polymer, Fe$_3$O$_4$@polymer, SnO$_2$@polymer~\cite{Li2012} and silica@polymer~\cite{Li2017}. Other compositions of eccentric core-shell particles were also developed, for instance, Ag@Ag$_2$S eccentric core-shell particles in nanometer size~\cite{Robinson2019} and TiO$_2$@SiO$_2$ non-concentric core-shell particles at the micron scale~\cite{Demirors2009}.

We explore Gaussian illumination on three types eccentric core-shell particles at the sub-micron scale: a gold nanoparticle coated with a silica shell (Au@SiO$_2$)~\cite{Chen2010, Li2014, Liu2017}, a titanium dioxide nanoparticle with a silica shell (TiO$_2$@SiO$_2$)~\cite{Demirors2009, Chemin2018} and a silicon dioxide particle with a titanium dioxide shell (SiO$_2$@TiO$_2$)~\cite{Rosales2020}. These are chosen as representative nanoparticles that may be fabricated with eccentric cores. We show that the optomechanical response of an eccentric spherical eccentric core-shell particle is richer than for a centro-symmetric symmetric particle, in particular due to the exertion of optical torques, as demonstrated in Sec.~3. This optomechanical response of such an eccentric core-shell particle can expand its remit in optical traps, namely inducing appreciable rotation rates in nanometre-sized spherical particles which opens up the prospect for nanorheology, which is the key motivation of this study. 

\section{Eccentric core-shell particle with Gaussian beam illumination}

\begin{figure*}[t]
\centering
\subfloat[]{ \includegraphics[width=0.35\textwidth]{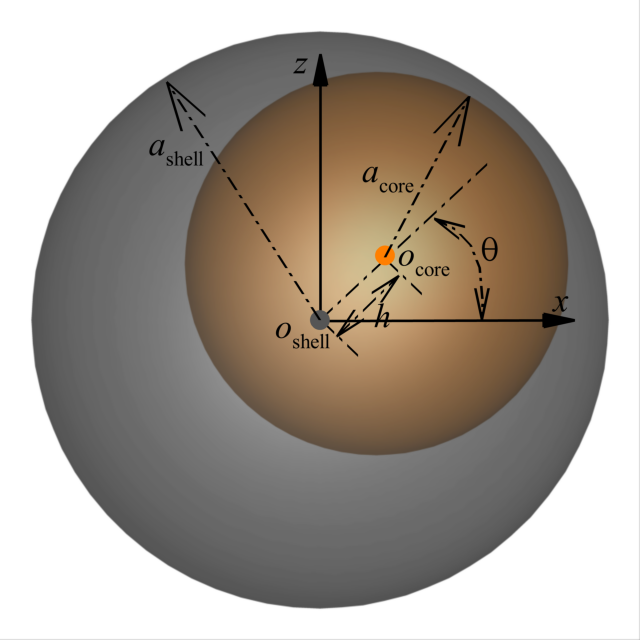}} \hspace{10mm}
\subfloat[]{ \includegraphics[width=0.45\textwidth]{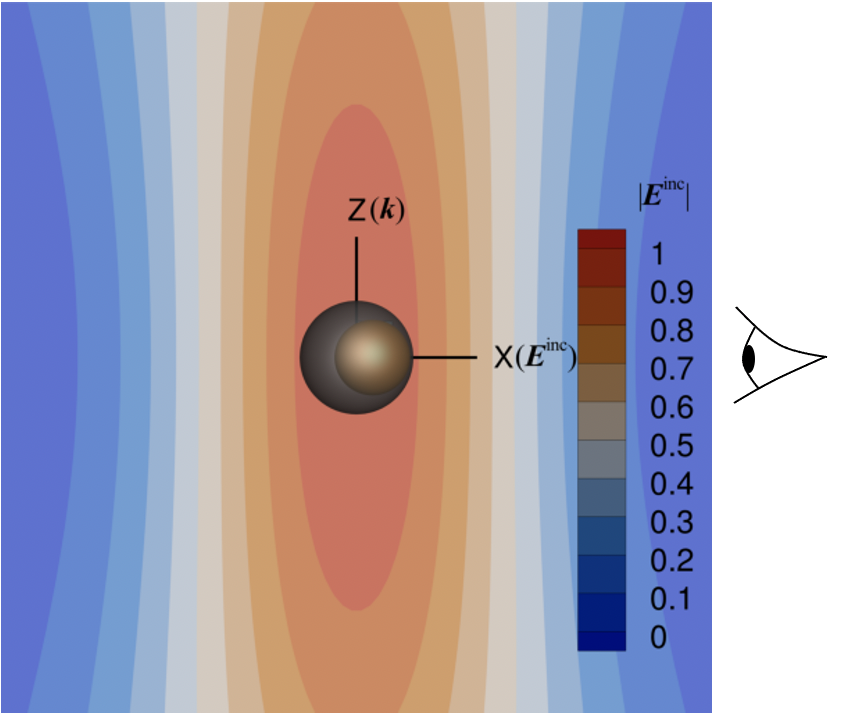}}
\caption{(a) Sketch of the geometric composition of the eccentric core-shell particle; (b) The electric field magnitude of the electric field of the Gaussian beam on the $xz$ plane. The beam propagates along the $z$-axis and its electric field is linearly polarised along the $x$-axis.} \label{Fig:fig1problem}
\end{figure*}

We consider particles where both the core and shell are spherical, and where the centers do not align, as shown in Fig.~\ref{Fig:fig1problem} (a). 

To induce optical effects, we firstly consider a trapping beam with Gaussian profile and linear polarisation. The focal position of the trapping beam is set at the origin, and the beam propagates along the $z$-axis with its electric field linearly polarising along the $x$-axis, as shown in Fig.~\ref{Fig:fig1problem} (b). 
The centre of the particle’s shell is fixed at the origin (the focal point of the beam), and the core-centre is varied on the $xz$ plane. The distance between the core-centre and shell-center is denoted as $h$, and the angle from the line connecting core-center and shell-center to the electric field polarisation direction ($x$ axis) is  $\theta$, so that when $\theta = 90^{\rmo}$ the centre line measured from the shell-centre to the core-centre is along the beam propagation direction and when $\theta = 270^{\rmo}$ it is opposite to the beam propagation direction.

When the beam is not highly focused, its Gaussian profile and the corresponding incoming electric and magnetic fields, $\boldsymbol{E}^{\rminc}$ and $\boldsymbol{H}^{\rminc}$, are well described by the approximate expressions given by Barton and Alexander~\cite{Barton1989, Barton1997}.  Assuming a linearly polarised single colour Gaussian beam propagating along the $z$ direction with its focal point locating at the origin in a source-free homogeneous medium with relative permittivity $\epsilon$ and permeability $\mu$, we may write the electric and magnetic fields as
\begin{subequations}\label{eq:EHinc}
    \begin{align}
        E_{x}^{\rminc} = & E_{0} \bigg \{ 1 + s^2(-\varrho^2 \vartheta^2 - \rmi \varrho^4 \vartheta^3 - 2\vartheta^2 \xi^2)  \nonumber \\
        & \qquad \quad +  s^4 \left[ 2\varrho^4\vartheta^4 + 3\rmi \varrho^6 \vartheta^5 - 0.5 \varrho^8 \vartheta^6 + (8 \varrho^2\vartheta^4 + 2 \rmi \varrho^4 \vartheta^5) \xi^2 \right] \bigg \}\psi_0 e^{\rmi kz}, 
    \end{align}
    \begin{align}
        E_{y}^{\rminc} = & E_{0} \left\{s^2(-2\vartheta^2 \xi \eta) + s^4 \left[ 8\varrho^2\vartheta^4 + 2\rmi \varrho^4 \vartheta^5) \xi \eta \right]
        \right\}\psi_0 e^{\rmi kz}, 
    \end{align}
    \begin{align}
        E_{z}^{\rminc} = & E_{0} \left\{s(-2\vartheta \xi) + s^3\left[ (6\varrho^2 \vartheta^3 +2 \rmi \varrho^4 \vartheta^4)\xi \right] \right. \nonumber \\
        & \qquad + \left. s^5 \left[ -20\varrho^4\vartheta^5 - 10\rmi \varrho^6 \vartheta^6 + \varrho^8 \vartheta^7   \right] \xi
        \right\}\psi_0 e^{\rmi kz}, 
    \end{align}
    \begin{align}
        H_{x}^{\rminc} = & \frac{k}{\mu_0 \mu \omega} E_{0} \left\{s^2(-2\vartheta^2 \xi \eta) + s^4 \left[ 8\varrho^2\vartheta^4 + 2\rmi \varrho^4 \vartheta^5) \xi \eta \right]
        \right\}\psi_0 e^{\rmi kz}, 
    \end{align}
    \begin{align}
        H_{y}^{\rminc} = & \frac{k}{\mu_0 \mu \omega} E_{0} \bigg \{ 1 + s^2(-\varrho^2 \vartheta^2 - \rmi \varrho^4 \vartheta^3 - 2\vartheta^2 \eta^2)  \nonumber \\
        & \qquad \quad +  s^4 \left[ 2\varrho^4\vartheta^4 + 3\rmi \varrho^6 \vartheta^5 - 0.5 \varrho^8 \vartheta^6 + (8 \varrho^2\vartheta^4 + 2 \rmi \varrho^4 \vartheta^5) \eta^2 \right] \bigg \}\psi_0 e^{\rmi kz}, 
    \end{align}
    \begin{align}
        H_{z}^{\rminc} = & \frac{k}{\mu_0 \mu \omega} E_{0} \left\{s(-2\vartheta \eta) + s^3\left[ (6\varrho^2 \vartheta^3 +2 \rmi \varrho^4 \vartheta^4)\eta \right] \right. \nonumber \\
        & \qquad + \left. s^5 \left[ -20\varrho^4\vartheta^5 - 10\rmi \varrho^6 \vartheta^6 + \varrho^8 \vartheta^7   \right] \eta
        \right\}\psi_0 e^{\rmi kz}, 
    \end{align}
In Eq.~(\ref{eq:EHinc}), $\omega$ is the angular frequency of the light beam, $k=\lambda/2\pi$ is the optical wave number with $\lambda$ the optical wavelength, $s = 1/(kw_0)$ with beam waist radius $w_0$, $\xi = x/w_0$, $\eta = y/w_0$, and $E_0$ is the electric field amplitude at the focal point of the beam ($x=y=z=0$) that is related to the beam power $P_0$ as
\end{subequations}
\begin{align}
    |E_{0}|^{2} = \frac{4P_0}{\pi w_0^{2} (1+s^2+1.5s^4)} \sqrt{\frac{\mu_0\mu}{\epsilon_0\epsilon}},
\end{align}
with $\epsilon_0$ and $\mu_0$ the free space permittivity and permeability, respectively. 
Also, functions $\varrho$, $\vartheta$ and $\psi_0$ are, respectively, defined as
\begin{align}
    \varrho=\sqrt{\xi^2+\eta^2}, \qquad \vartheta = \frac{kw_0^2}{2z - \rmi kw_0^2}, \qquad \psi_0 = -\rmi \vartheta e^{\rmi \varrho^2 \vartheta}.
\end{align}
If the trapping beam with Gaussian profile is not linearly polarised but circularly polarised, the electric and magnetic fields can be also described based on Eq.~(\ref{eq:EHinc})~\cite{Barton1997}, as detailed in Sec.~SI-2 in the supporting information~\cite{Sun2020SI}.

The scattered fields, $\boldsymbol{E}^{\rmsc}$ and $\boldsymbol{H}^{\rmsc}$, and the transmitted fields in the particles are calculated by using the robust field only surface integral method~\citep{Sun2017-RMF, Klaseboer2017-FOI, Sun2020-PEC, Sun2020-DIEL}, which is ideally suited to calculating the electric and magnetic fields efficiently and accurately on particle surfaces.

By obtaining the surface electric and magnetic fields accurately, we then calculate the optical force by integrating the time-average Maxwell’s stress tensor over the surface of the shell, $S_{\text{shell}}$, as
\begin{align}\label{eq:force_tot}
    F_{i} = &\int_{S_{\text{shell}}} \frac{1}{2} \Big\{\text{Real} \left[ (D_{i}E_{j}^{*} + E_{i}D_{j}^{*} + B_{i}H_{j}^{*} + H_{i}B_{j}^{*}) n_{j} - (D_{j}E_{j}^{*} + B_{j}H_{j}^{*} )n_{i}  \right] \Big\} \rmd S. 
\end{align}
In Eq.~(\ref{eq:force_tot}), subscript $i =1,2,3$ is the $i$th component of the vector field, subscript $j =1,2,3$ is the $j$th component of the vector field, superscript $^*$ indicates the conjugate of the field. Also, $F_{i}$ is the optical force, $n_i$ is the unit normal vector on surface $S_{\text{shell}}$, $E_{i}$ is the $i$th component of electric field, $D_{i}$ is the $i$th component of electric displacement, $H_{i}$ is the $i$th component of magnetizing field, and $B_{i}$ is the $i$th component of magnetic field. In Eqs.~(\ref{eq:force_tot}) and~(\ref{eq:torque_tot}), the Einstein notation is used, and the 1st, 2nd and 3rd components correspond to the $x$, $y$ and $z$ components of field, respectively.

The optical torque can be calculated by
\begin{align}\label{eq:torque_tot}
    N_{i} = & \int_{S_{\text{shell}}} \varepsilon_{ijk} r^{c}_{j} \frac{1}{2} \Big\{\text{Real}  \left[ (D_{k}E_{l}^{*} + E_{k}D_{l}^{*} + B_{k}H_{l}^{*} + H_{k}B_{l}^{*}) n_{l} - (D_{l}E_{l}^{*} + B_{l}H_{l}^{*} )n_{k}  \right] \Big\} \rmd S.
\end{align}
where subscript $l =1,2,3$ is the $l$th component of the vector field, subscript $k =1,2,3$ is the $k$th component of the vector field, $\varepsilon_{ijk}$ is the Levi-Civita symbol, and $r^{c}_{j}$ is the location vector of a point on the shell surface to the centre of mass of the core-shell particle. The centre of mass of the core-shell particle is on the centre line between the centre of the core and that of the shell which distance to the centre of the shell, $h_c$, can be calculated by $h_c = h(\rho_{\text{core}} - \rho_{\text{shell}})/(\rho_{\text{core}} + \rho_{\text{shell}})$ where $\rho_{\text{core}} $ is the density of the core and $\rho_{\text{shell}}$ is the density of the shell.

\section{Results and discussion}

We study the optomechanical response of three types of eccentric core-shell particles held with a trapping beam with  a Gaussian profile, namely Au@SiO$_2$, TiO$_2$@SiO$_2$ and SiO$_2$@TiO$_2$. The densities, refractive indices and extinction coefficients of Au, SiO$_2$ and TiO$_2$ are listed in Table~\ref{tab:nk}. For our purposes, we assume that the particle has been trapped with the centre of the shell located at the focal point of the beam. We can therefore concentrate on how the orientation of the core particle will affect the optical trapping of the eccentric core-shell particle. In this section, we focus on the optical torque acting on the nanoscale eccentric spherical core-shell particle for the case of both a linearly polarised and a circularly polarised Gaussian beam. The case of linear polarisation may lead to an optical torque wrench~\cite{Santybayeva2016} with the particle aligning with the field or continuously rotating as the polarisation itself is rotated. For a circularly polarised trapping beam we would see continuous rotation. For the cases of continuous rotation, the particle may reach a terminal angular velocity once the applied optical torque matches the rotational Stokes drag. The optical trapping forces of the eccentric core-shell spherical particles are discussed in detail in Sec.~SI-1 in the Supporting Information~\cite{Sun2020SI}.

\begin{table}[t]
\centering
\caption{ \bf Densities, refractive indices $n$ and extinction coefficients $k$ of Au~\cite{Raki1998}, SiO$_2$~\cite{Malitson1965} and TiO$_2$~\cite{Bodurov2016} }
\begin{tabular}{l | c | c | c | c | c }
\hline
& Density (g/cm$^3$) & \multicolumn{4}{c}{$n + \rmi k$} \\
\cline{3-6} 
& & $\lambda$=532 nm & $\lambda$=775 nm & $\lambda$=840 nm & $\lambda$=1064 nm  \\
\hline
Au      & 19.30 & 0.54+$\rmi$2.14 & 0.18+$\rmi$4.51 & 0.20+$\rmi$5.02 & 0.31+$\rmi$6.63 \\
SiO$_2$ & 2.50 & 1.46            & 1.45            & 1.45            & 1.45            \\
TiO$_2$ & 4.23 & 2.17            & 2.10            & 2.09            & 2.07            \\
\hline
\end{tabular}
\label{tab:nk}
\end{table}

\begin{figure*}[!ht]
\centering{}
\subfloat[Au@SiO$_2$, $a_{\text{Au}}$=60nm, $a_{\text{SiO}_{2}}$=90nm, $h$=25nm]{ \includegraphics[width=0.32\textwidth]{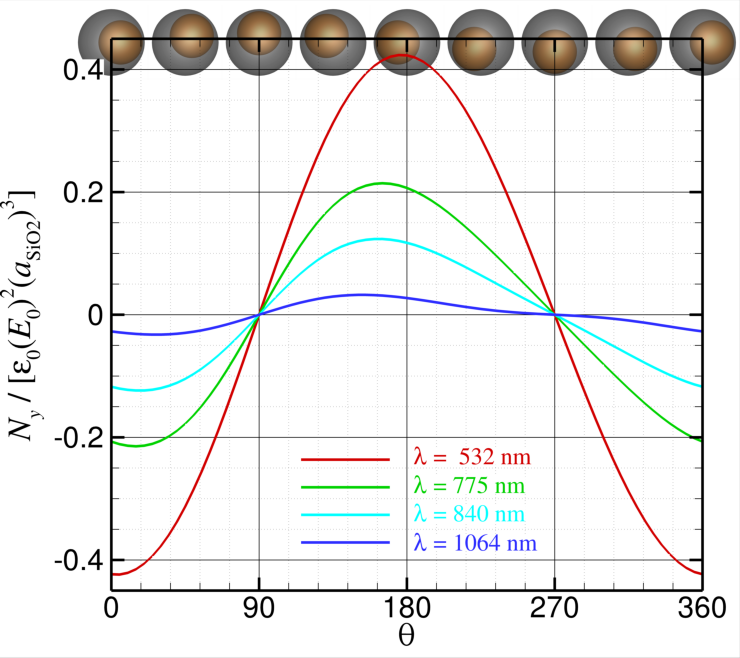}}
\subfloat[TiO$_2$@SiO$_2$, $a_{\text{TiO}_{2}}$=60nm, $a_{\text{SiO}_{2}}$=90nm, $h$=25nm]{ \includegraphics[width=0.32\textwidth]{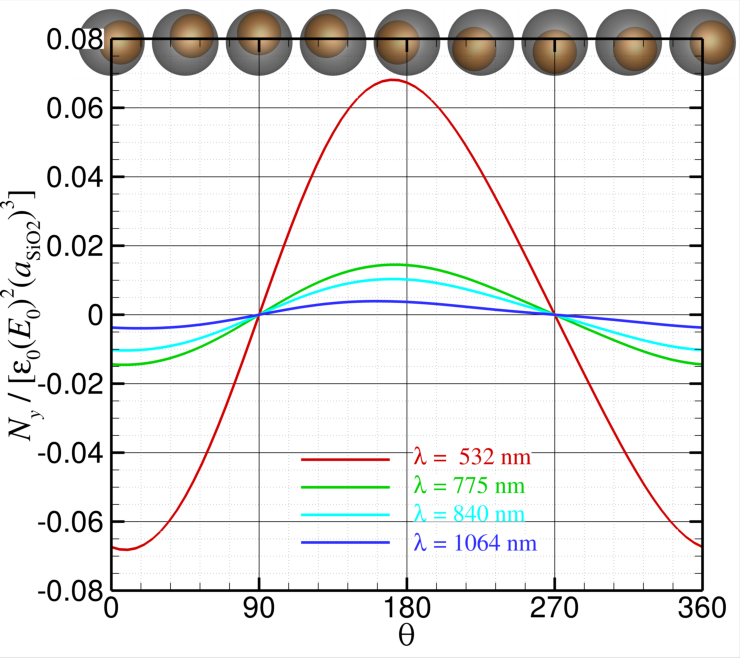}}
\subfloat[SiO$_2$@TiO$_2$, $a_{\text{SiO}_{2}}$=60nm, $a_{\text{TiO}_{2}}$=90nm, $h$=25nm]{ \includegraphics[width=0.32\textwidth]{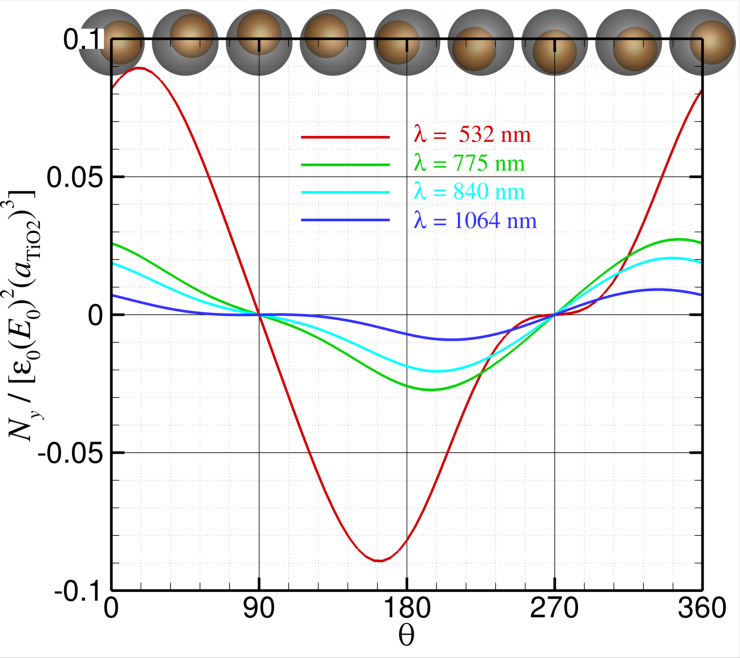}} \\
\subfloat[Au@SiO$_2$, $a_{\text{Au}}$=60nm, $a_{\text{SiO}_{2}}$=90nm, $h=10$ nm]{ \includegraphics[width=0.32\textwidth]{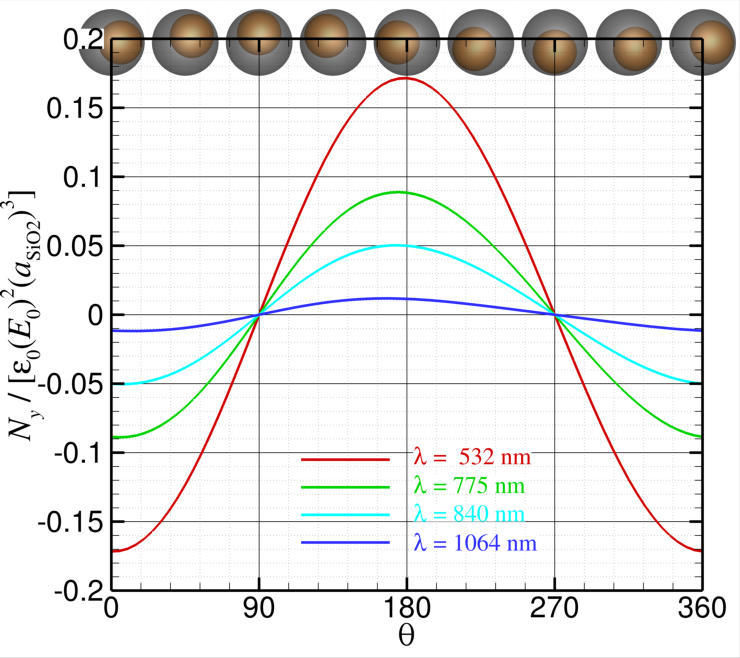}}
\subfloat[TiO$_2$@SiO$_2$, $a_{\text{TiO}_{2}}$=60nm, $a_{\text{SiO}_{2}}$=90nm, $h=10$ nm]{ \includegraphics[width=0.32\textwidth]{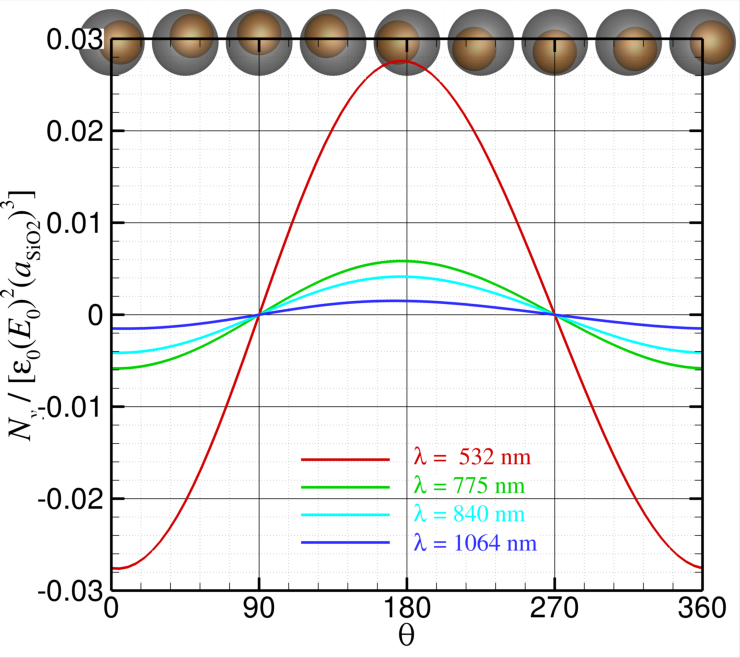}}
\subfloat[SiO$_2$@TiO$_2$, $a_{\text{SiO}_{2}}$=60nm, $a_{\text{TiO}_{2}}$=90nm, $h=10$ nm]{ \includegraphics[width=0.32\textwidth]{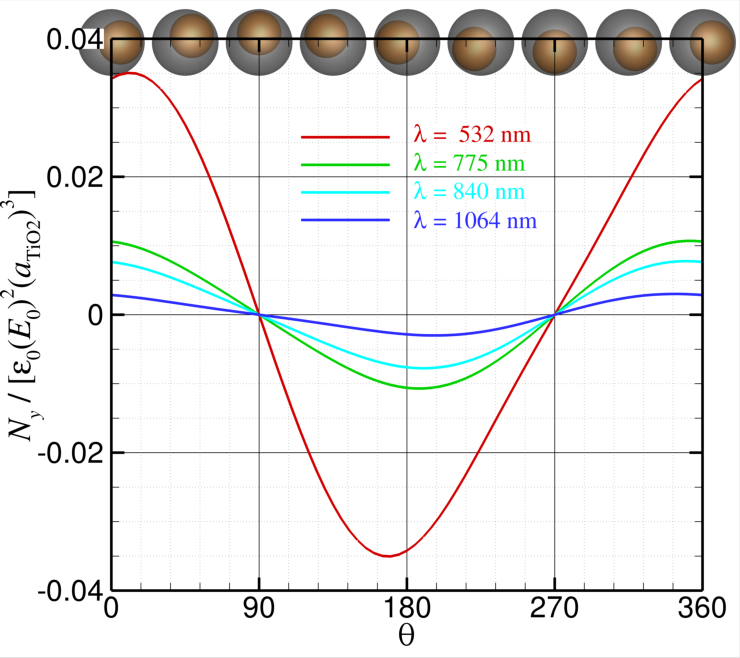}} \\
\subfloat[Au@SiO$_2$, $a_{\text{Au}}$=30nm, $a_{\text{SiO}_{2}}$=45nm, $h=10$ nm]{ \includegraphics[width=0.32\textwidth]{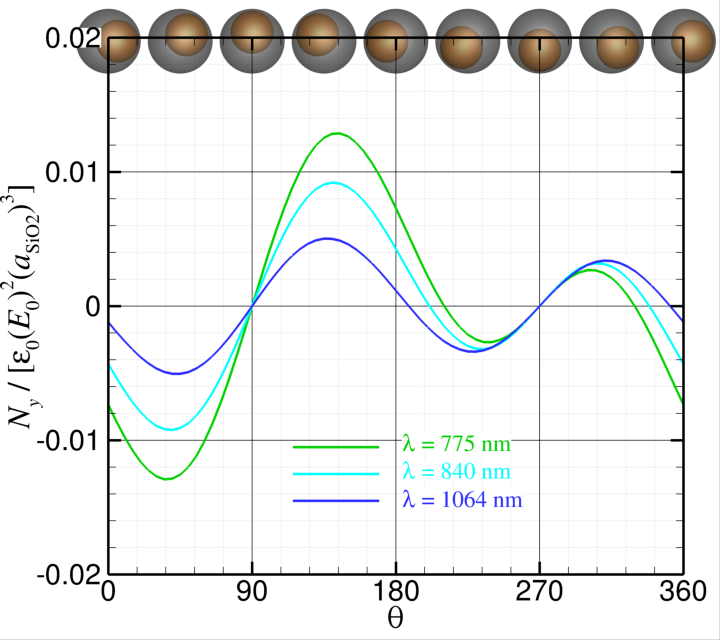}}
\subfloat[TiO$_2$@SiO$_2$, $a_{\text{TiO}_{2}}$=30nm, $a_{\text{SiO}_{2}}$=45nm, $h=10$ nm]{ \includegraphics[width=0.32\textwidth]{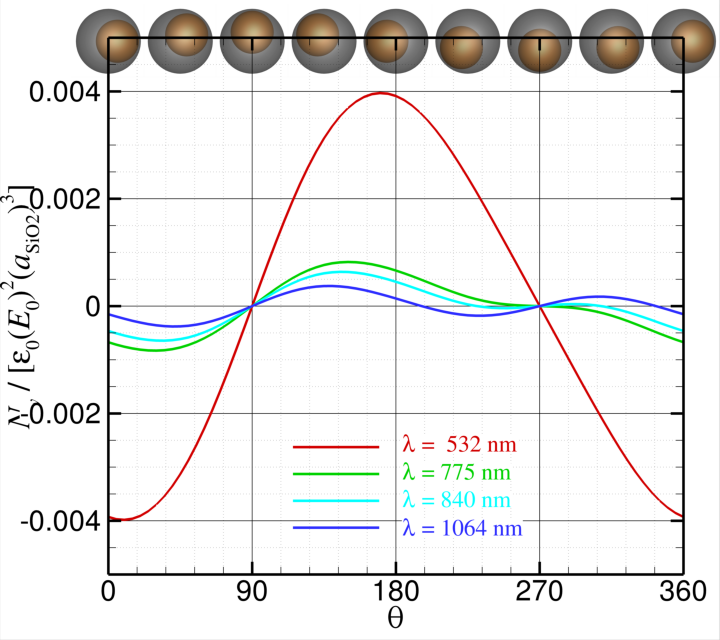}}
\subfloat[SiO$_2$@TiO$_2$, $a_{\text{SiO}_{2}}$=30nm, $a_{\text{TiO}_{2}}$=45nm, $h=10$ nm]{ \includegraphics[width=0.32\textwidth]{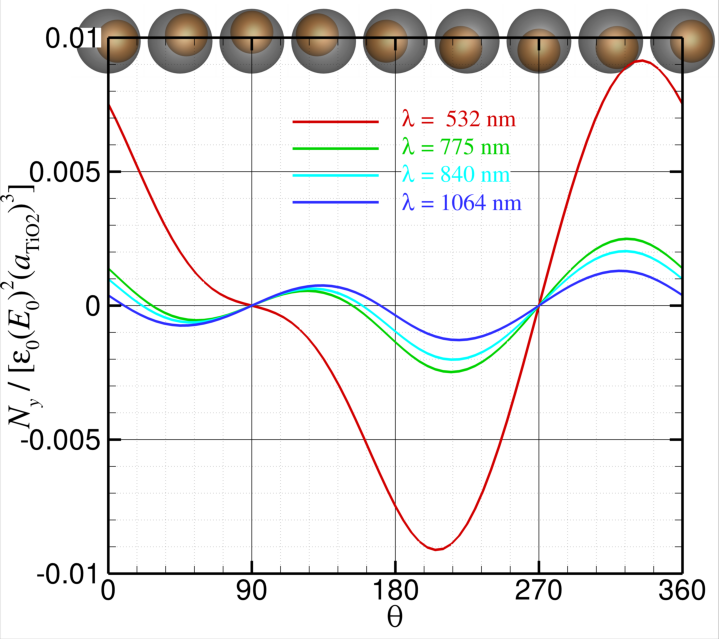}} 
\caption{Optical torque acting on three types of eccentric core-shell particles in water with $n_{\text{water}}=1.33$ under the linearly polarised Gaussian beam illumination with different wavelength $\lambda$ and beam waist radius fixed at $w_0=1$ $\mu$m. The Au@SiO$_2$ particles show the most pronounced effects due to the larger refractive index contrast, with approximately one order of magnitude greater optical torque than for the all dielectric particles.  Interestingly both TiO$_2$@SiO$_2$ and SiO$_2$@TiO$_2$ particles show comparable optical torques with approximately inverted responses as a function of particle orientation.
}  \label{Fig:torque}
\end{figure*}

In Fig.~\ref{Fig:torque}, we show the optical torques acting on three types of eccentric core-shell particles embedded in water under a linearly polarised Gaussian beam for a few different wavelength values, with a fixed beam waist radius $w_0=1~\mu$m. We chose readily available standard wavelengths of $\lambda=532$~nm, $\lambda=775$~nm, $\lambda=840$~nm, and $\lambda=1064$~nm.
Also, these wavelengths avoid the consequence of surface plasmon resonance for Au@SiO$2$ in water with $a_{\text{SiO}_{2}}=90$ nm and $a_{\text{Au}}=60$ nm that happens at wavelength around 640 nm.

We can see, from Fig.~\ref{Fig:torque}, that as the wavelength becomes longer, the magnitude of the optical torque decreases. Also, the optical torque exerted on Au@SiO$_2$ is nearly one order of magnitude higher than that on TiO$_2$@SiO$_2$ or SiO$_2$@TiO$_2$. This is mainly because the difference of the refractive index between Au and SiO$_2$ is higher than that between TiO$_2$ and SiO$_2$ (see  Table~\ref{tab:nk}). Also, the trend of optical torque along with the orientation of the core particle, $\theta$ on TiO$_2$@SiO$_2$ with $n_{\text{core}}>n_{\text{shell}}>n_{\text{medium}}$ is opposite to that on SiO$_2$@TiO$_2$ with $n_{\text{shell}}>n_{\text{core}}>n_{\text{medium}}$, as shown in the second and third columns of Fig.~\ref{Fig:torque}. This is another indicator of how the material refractive index can affect the optomechanical response of an eccentric core-shell particle. Similar to the optical force perpendicular to the beam direction, the maximum optical torques appear at $\theta=0^{\rmo}$ and $\theta=180^{\rmo}$ for the core-shell particle with shell radius 90 nm and core radius 60 nm. For Au@SiO$_2$ and TiO$_2$@SiO$_2$ eccentric core-shell particles, when $\theta=0^{\rmo}$, the optical torque will rotate the eccentric particle counter-clockwise if we define that the beam propagation direction is pointing at 12 o'clock, and when $\theta=180^{\rmo}$ the optical torque will rotate them clockwise. However, for SiO$_2$@TiO$_2$, the direction of the optical torque flips relative to Au@SiO$_2$ or TiO$_2$@SiO$_2$.

As expected, the optical torque increases as the core particle is positioned further away from the centre of the shell (i.e. the system is further from centrosymmetry). From the synthesis point of view, moderate asymmetry with small $h$ may be easier to achieve. In the first two rows of Fig~\ref{Fig:torque}, we demonstrated the optical torques on Au@SiO$_2$, TiO$_2$@SiO$_2$ and SiO$_2$@TiO$_2$ with $h=25$ nm (top row) for high asymmetry and $h=10$ nm (middle row) for moderate asymmetry when the shell radius is 90~nm and the core radius 60~nm. The optical torques with $h=25$~nm are more than twice those with $h=10$ nm. By using the Stokes drag of a sphere in fluid, $N_y = 8\pi \mu a_{\text{shell}}^{3} \Omega$ where $\mu$ is the viscosity of water and $\Omega$ is the particle rotation frequency, we can estimate the possible largest rotation frequency of the eccentric core-shell particles, when held by a linearly polarised Gaussian beam, for different wavelengths, each with a fixed beam power of $P_0 = $ 20 mW. Such a rotation can be achieved by using the optical torque wrench~\cite{Santybayeva2016}. We observe that even for small particle sizes where the radius is 90 nm (diameter of 180 nm), the Au@SiO$_2$ eccentric core-shell particle can be rotated at frequencies of a few hundred Hz with 532~nm. For $\lambda= 1064$~nm, the rotation frequency of the Au@SiO$_2$ can reach up to 61 Hz with $h=25$ nm and 22 Hz with $h=10$ nm. As to TiO$_2$@SiO$_2$ and SiO$_2$@TiO$_2$ eccentric core-shell particles, the potential rotation frequencies can still range from a few Hz to 50~Hz. Our results are summarised in Table~\ref{tab:frequency}.

\begin{table*}[t]
\centering
\caption{\bf Potential maximum rotation frequency $\Omega$ (in Hz) by using the optical torque wrench~\cite{Santybayeva2016} for three types of eccentric spherical core-shell particles suspended in water ($n=1.33$) and illuminated with a linearly polarised Gaussian beam at different wavelengths, $\lambda$. The beam waist radius is 1$\mu$m and beam power is $P_0 = $ 20 mW in all cases}
\resizebox{1.0\textwidth}{!}{
\begin{tabular}{l | c | c | c | c | c | c | c | c | c}
\hline
& \multicolumn{3}{c|}{Au@SiO$_2$} & \multicolumn{3}{c|}{TiO$_2$@SiO$_2$} & \multicolumn{3}{c}{SiO$_2$@TiO$_2$} \\
\cline{2-10}
& \multicolumn{2}{c|}{$a_\text{shell}$=90 nm} & $a_\text{shell}$=45 nm & \multicolumn{2}{c|}{$a_\text{shell}$=90 nm} &
$a_\text{shell}$=45 nm&
\multicolumn{2}{c|}{$a_\text{shell}$=90 nm} &
$a_\text{shell}$=45 nm \\
& \multicolumn{2}{c|}{$a_\text{core}$=60 nm} & $a_\text{core}$=30 nm & \multicolumn{2}{c|}{$a_\text{core}$=60 nm} &
$a_\text{core}$=30 nm &
\multicolumn{2}{c|}{$a_\text{core}$=60 nm} &
 $a_\text{core}$=30 nm \\
\cline{2-10}
& $h$=25 nm & $h$=10 nm  & $h$=10 nm & $h$=25 nm &  $h$=10 nm  &  $h$=10 nm  & $h$=25 nm & $h$=10 nm  &  $h$=10 nm \\
\hline
$\lambda$=532 nm  & 807 & 327 & -  & 130 & 53 & 8 & 170 & 67 & 17 \\
$\lambda$=775 nm  & 407 & 169 & 24 & 28  & 11 & 2 & 52  & 20 & 5 \\
$\lambda$=840 nm  & 234 & 95  & 17 & 20  & 8  & 1 & 39  & 15 & 4 \\
$\lambda$=1064 nm & 61  & 22  & 9  & 7   & 3  & 1 & 17  & 6  & 2 \\
\hline
\end{tabular}
}
\label{tab:frequency}
\end{table*}

If the particle size is reduced by half, for instance $a_{\text{shell}}=45$~nm and $a_{\text{core}}=30$~nm, the optical torques on the eccentric core-shell particles are still observable, as shown in Fig.~\ref{Fig:torque}(g,h,i). Note that in Fig.~\ref{Fig:torque}(g), we did not show the optical torque with $\lambda=532$ nm since that wavelength is close to the surface plasmon resonance wavelength that is around 550 nm for Au@SiO$_2$ core-shell particle in water when $a_{\text{SiO}_{2}}=45$ nm and $a_{\text{Au}}=30$ nm. Based on the Stokes drag, the rotation frequency of such a small particle with diameter less than 100~nm can be a few Hz when the optical torque wrench beam power is only 20~mW, as displayed in Table~\ref{tab:frequency}. This result is significant for practical applications, in particular biological applications, as particle sizes below 100 nm are potentially more bio-compatible than larger particles. Furthermore, using such low powers for optical rotation should reduce potential issues associated with photo toxicity. Also, we would like to emphasise that in practice, the optical trapping location of the particle is not at the focal point of beam in most cases due to the field gradient. We consider a few cases when the geometrical centre of the core-shell particle is located at different positions relative to be focus of beam along the beam propagation direction, denoted as $(0,0,z^c_{\text{shell}})$. As shown in detail in Sec.~SI-1.4 in the Supporting Information~\cite{Sun2020SI}, we find that the magnitude of the trapping force $F_z$ changes significantly when $z^c_{\text{shell}}$ varies from $z^c_{\text{shell}}=0$ $\mu$m to $z^c_{\text{shell}}=1$ $\mu$m, $z^c_{\text{shell}}=2$ $\mu$m and $z^c_{\text{shell}}=4$ $\mu$m while the magnitude of the optical torque $N_y$ does not show an obvious variation. This indicates that we can observe the optical rotation of the eccentric core-shell in nanoscale in practical experiments. This indicates that the optical rotation of nanoscale eccentric core-shell particles should be observable in realistic experimental scenarios, performed with current technology.

\begin{figure*}[t]
\centering{}
\subfloat[TiO$_2$@SiO$_2$]{ \includegraphics[width=0.32\textwidth]{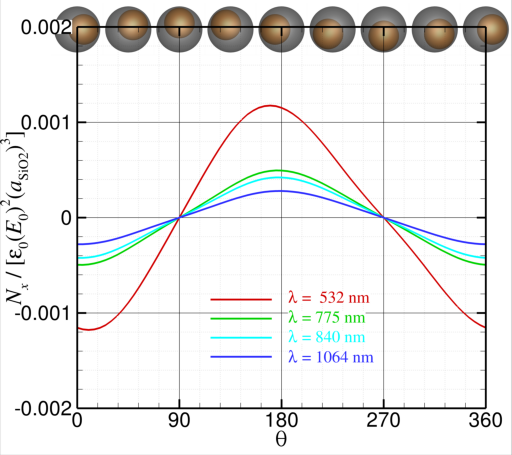}}
\subfloat[TiO$_2$@SiO$_2$]{ \includegraphics[width=0.32\textwidth]{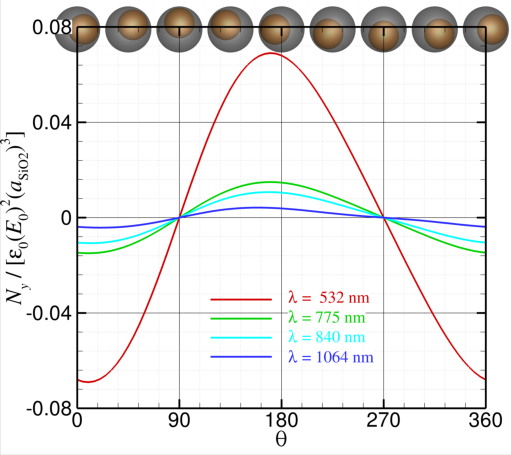}}
\subfloat[TiO$_2$@SiO$_2$]{ \includegraphics[width=0.32\textwidth]{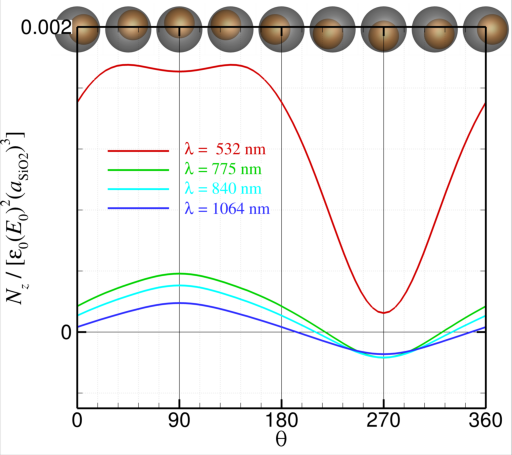}} \\
\subfloat[SiO$_2$@TiO$_2$]{ \includegraphics[width=0.32\textwidth]{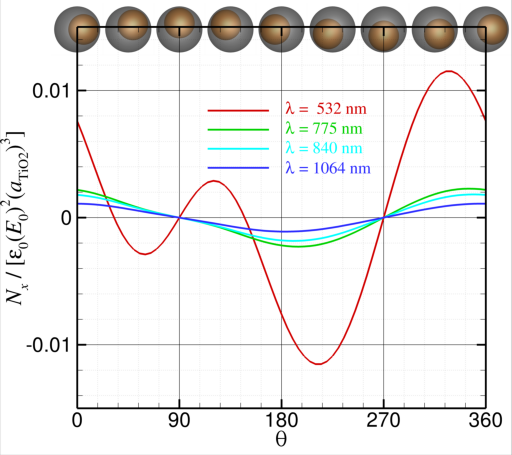}}
\subfloat[SiO$_2$@TiO$_2$]{ \includegraphics[width=0.32\textwidth]{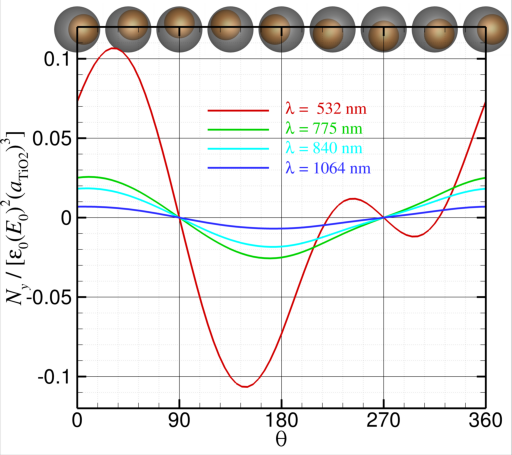}}
\subfloat[SiO$_2$@TiO$_2$]{ \includegraphics[width=0.32\textwidth]{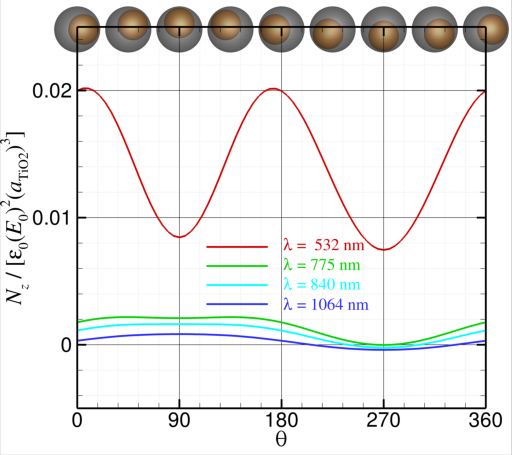}}
\caption{Optical torque on (a-c) an eccentric TiO$_2$@SiO$_2$ core-shell particle and (d-f) an eccentric SiO$_2$@TiO$_2$ in water ($n_{\text{medium}} = 1.33$) illuminated by a circularly polarised Gaussian beam with waist radius of $w_0=1$ $\mu$m. The geometric features of the eccentric core-shell particle are $a_{\text{shell}}=90$ nm, $a_{\text{core}}=60$ nm and $h=25$ nm. The magnitude of torque $N_y$ that is introduced by the asymmetry of the eccentric core-shell particle is much higher than that of torque $N_z$ which is induced by the circular polarisation of the beam.
}  \label{Fig:Ncp_TiO2SiO2}
\end{figure*}

To date, the most popular demonstration of optical rotation of small particle is to use a circularly polarised Gaussian beam to rotate a birefringent particle at the micro-scale since a birefringent particle has two nonparallel optical paths with different refractive indices.  We calculated the optical torque under the Gaussian illumination with circular polarisation on an eccentric spherical core-shell particle at nanoscale with a dielectric core embedded in a dielectric shell, such as TiO$_2$@SiO$_2$ and SiO$_2$@TiO$_2$. The mathematical description of the illumination beam is given in Sec.~SI-2 of the Supporting Information~\cite{Sun2020SI}. Fig.~\ref{Fig:Ncp_TiO2SiO2} demonstrates how the optical torques acting on an eccentric TiO$_2$@SiO$_2$ and an eccentric SiO$_2$@TiO$_2$ core-shell particle at nanoscale emerged in water ($n_{\text{medium}}=1.33$) change with respect to the variation of the orientation, $\theta$, when the shell is located at the focus of the Gaussian beam with circular polarisation. In this case, when $\theta$ changes, the optical torques along all three directions, $N_x$, $N_y$ and $N_z$, appear in which $N_x$ and $N_y$ are induced by the asymmetry of the eccentric spherical core-shell particle which $N_z$ is mainly due to the circular polarisation of the beam.  Since in our calculations, we set variation of the geometrical feature (asymmetry) of the eccentric core-shell particle to happen in the $xz$ plane, $N_y$ is the dominating torque relative to $N_x$ as shown in the first and the second column of Fig.~\ref{Fig:Ncp_TiO2SiO2}. Relative to Fig.~\ref{Fig:torque}, we can see that the amplitude of $N_y$ under the circularly polarised Gaussian beam is in the same order as that under the linearly polarised Gaussian beam, as shown in the second column of Fig.~\ref{Fig:Ncp_TiO2SiO2}. Nevertheless, we find that the magnitude of torque $N_z$ that is induced by the circular polarisation of the beam is much weaker when compared to the magnitude of torque $N_y$ that is introduced by the asymmetry of the eccentric core-shell particle, as presented in second and third columns of Fig.~\ref{Fig:Ncp_TiO2SiO2}. As such, we can conclude that the idea to exploit the asymmetry of eccentric core-shell particle for optical rotation is more robust and general when compared to solely using the circular polarisation of the trapping beam.  

\begin{figure*}[t]
\centering{}
\subfloat[Au@SiO$_2$]{ \includegraphics[width=0.3\textwidth]{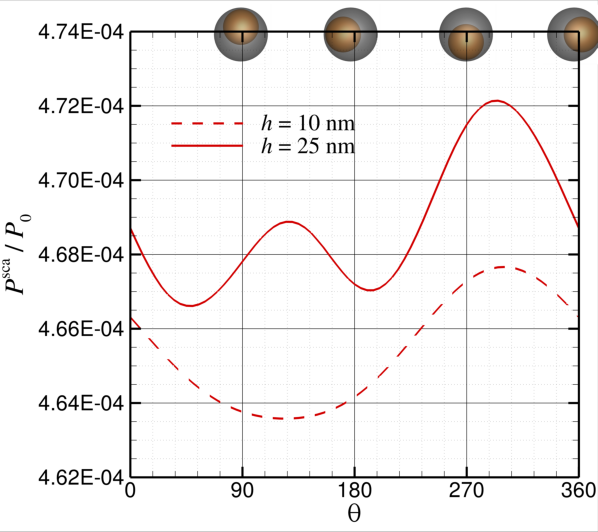}}
\subfloat[TiO$_2$@SiO$_2$]{ \includegraphics[width=0.3\textwidth]{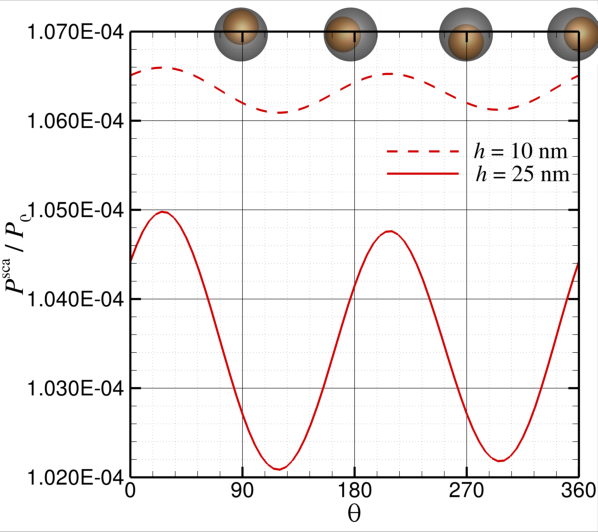}}
\subfloat[SiO$_2$@TiO$_2$]{ \includegraphics[width=0.3\textwidth]{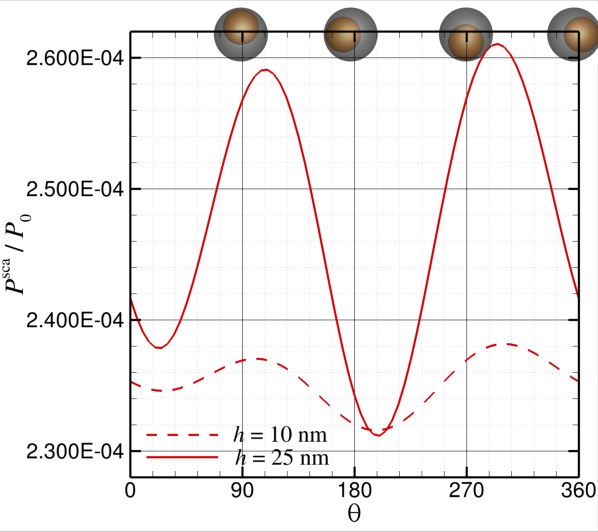}}
\caption{The scattered power along with the orientation of the core when $\lambda = 532$ nm that is collected by an objective facing the beam polarisation direction. The objective is a circular plane with radius of 3.75 mm which centre is located at (0.13 mm, 0, 0). In all cases, the power in the scattered beam varies by of order $1\%$, as a function of particle orientation, which is readily detectable.
}  \label{Fig:SnPower}
\end{figure*}

To confirm that the rotation of an eccentric core-shell particle is an observable phenomenon, we studied the scattered light from the particle, $P^{\rmsc}$ from the side of the linearly polarised light beam. The scattered light is collected by a circular objective with radius of 3.75~mm with centre located at (0.13~mm, 0, 0), as shown in Fig.~\ref{Fig:fig1problem}b (indicated by the eye). The power obtained by the objective can be calculated by integrating the time-average Poynting vector over the objective as $P^{\rmsc} = \int_{S_{\text{objective}}} \frac{1}{2} \{\text{Real}  [ \boldsymbol{E}^{\rmsc} \times (\boldsymbol{H}^{\rmsc})^{*} ] \} \rmd {\boldsymbol{S}}$. Fig.~\ref{Fig:SnPower} presents the light power (photon counts) at the far field perpendicular to the beam propagation when an eccentric core-shell particle with $a_{\text{shell}}=90$~nm and $a_{\text{core}}=60$~nm trapped by a Gaussian beam with wavelength of 532 nm and beam waist radius of 1~$\mu$m. From this figure, we can see that the scattered light power, $P^{\rmsc}$ varies along with the orientation of the core, which can be used to monitor the rotation of the eccentric core-shell particles in practical experiments.

\section{Conclusion}

We numerically studied the optomechanical response of an eccentric core-shell particle with sub-micron size trapped by a Gaussian beam with fixed beam waist radius $w_0=1$ $\mu$m (more detailed results can found in the supplementary material~\cite{Sun2020SI}). We considered three types of core-shell particles: Au@SiO$_2$, TiO$_2$@SiO$_2$ and SiO$_2$@TiO$_2$ together with different optical trapping light beam wavelengths: $\lambda=532$ nm, $\lambda = 775$ nm, $\lambda=840$ nm and $\lambda=1064$ nm. One exciting observation is that such an eccentric spherical core-shell particle with diameter of 180 nm (Au@SiO$_2$) can be rotated at over 800 Hz for a moderate illumination power of 20 mW. For the same power, a 90 nm diameter particle (SiO$_2$@TiO$_2$) can be rotated at over 10 Hz. Optical rotation of such a small particle with such moderate light powers is not achievable with currently used birefringent particles. This indicates that the use of eccentric sub-micron scale core-shell particles will find use in fundamental and applied studies in optical trapping where researchers seek rotation of nanometric sized objects. This could include the areas of levitated optomechanics and biophotonics, where the particle may perform as an optical torque wrench or enable the local measurement of nano-viscosity in complex fluids, a path way to \textit{nanorheology}. This is highly relevant to gain an understanding of how cells respond to stimuli from their surrounding environment.

\begin{acknowledgement}

QS and ADG acknowledge support of the Australian Research Council Centre of Excellence for Nanoscale BioPhotonics (CNBP) (Grant No. CE140100003). KD acknowledges the UK Engineering and Physical Sciences Research Council (Grant EP/P030017/1). QS acknowledges the support of an Australian Research Council Discovery Early Career Researcher Award (Grant No. DE150100169), and ADG acknowledges the support of an Australian Research Council Future Fellowship (Grant No. FT160100357). This research was undertaken with the assistance of resources from the National Computational Infrastructure (NCI Australia), an NCRIS enabled capability supported by the Australian Government (Grant No. LE160100051). 

\end{acknowledgement}

\begin{suppinfo}


The following files are available free of charge.
\begin{itemize}
  \item Supporting information for optical forces and torques on eccentric nanoscale core-shell particles~\cite{Sun2020SI}
\end{itemize}

\end{suppinfo}

\bibliography{OTCSref}

\end{document}


\maketitle

Number of pages: 15

Number of figures: 10

Number of tables: 1

\clearpage

\newpage

\section{Optomechanical response of an eccentric spherical core-shell particle under the Gaussian illumination with linear polarisation}

Here we provide the detailed optomechanical response of an eccentric spherical core-shell particle under the Gaussian illumination with linear polarisation, as described in Sec. 2 of the main text. The asymmetry of the particle is introduced by the displacement between the core-centre and shell centre. The beam waist radius is set as $w_0=1$ $\mu$m when the beam propagates along $z$-axis and polarises along $x$-axis. The distance between the core-centre and shell-center is $h$, and the angle from the line connecting core-center and shell-center to the $x$ axis is  $\theta$, so that when $\theta = 90^{\rmo}$ the centre line measured from the shell-centre to the core-centre is along the beam propagation direction.

\begin{figure*}
\centering{}
\subfloat[$\theta = 0^{\rmo}$]{ \includegraphics[width=0.32\textwidth]{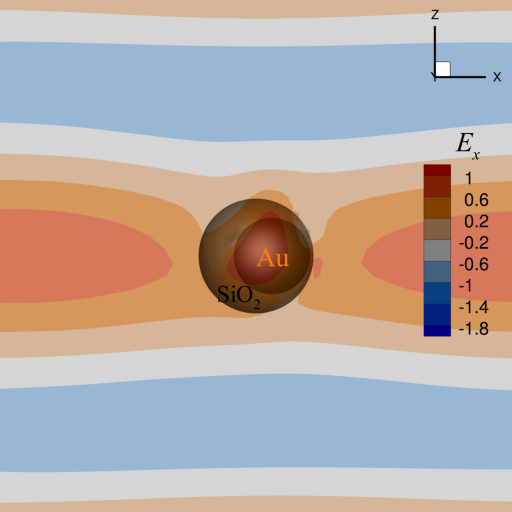}}
\subfloat[$\theta = 45^{\rmo}$]{ \includegraphics[width=0.32\textwidth]{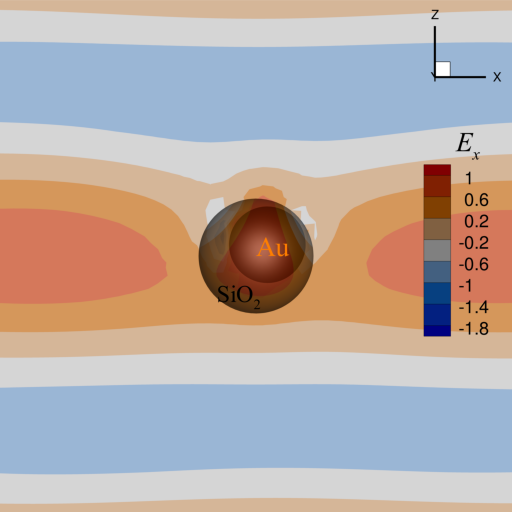}}
\subfloat[$\theta = 315^{\rmo}$]{ \includegraphics[width=0.32\textwidth]{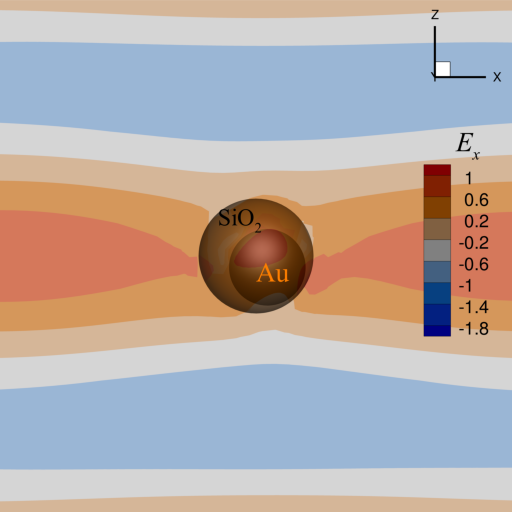}} \\
\subfloat[$\theta = 0^{\rmo}$]{ \includegraphics[width=0.32\textwidth]{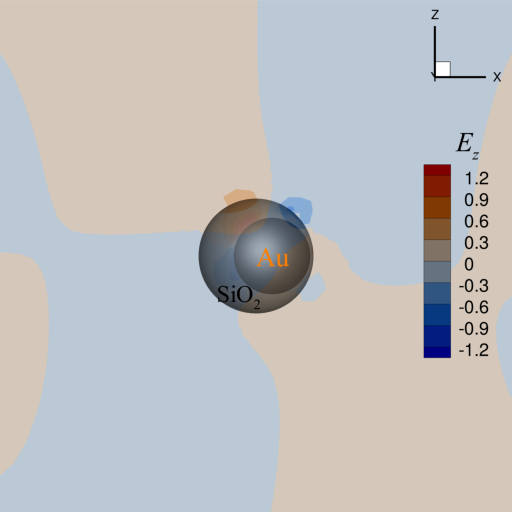}}
\subfloat[$\theta = 45^{\rmo}$]{ \includegraphics[width=0.32\textwidth]{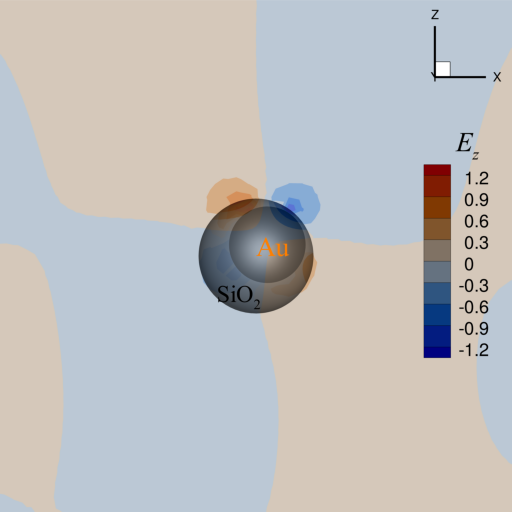}}
\subfloat[$\theta = 315^{\rmo}$]{ \includegraphics[width=0.32\textwidth]{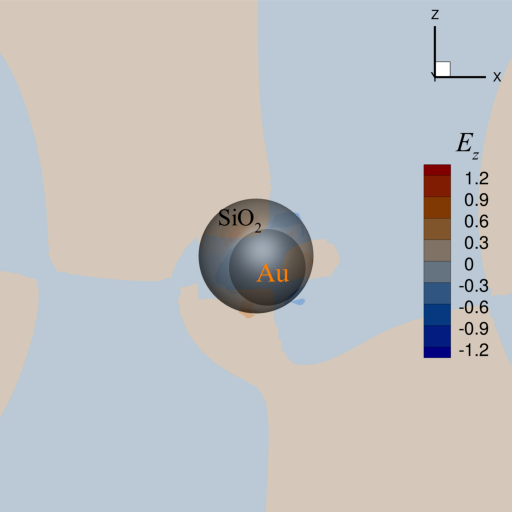}} \\
\subfloat[$\theta = 0^{\rmo}$]{ \includegraphics[width=0.32\textwidth]{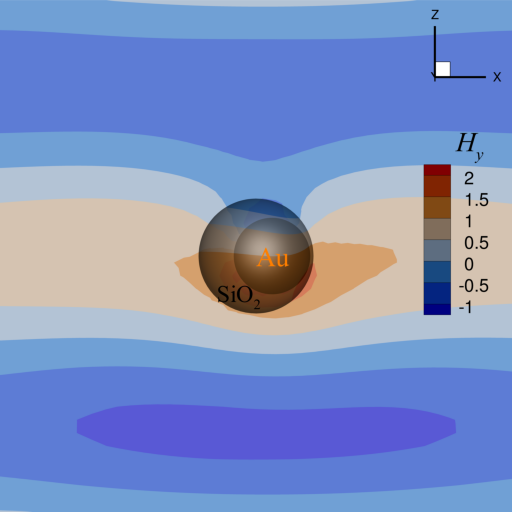}}
\subfloat[$\theta = 45^{\rmo}$]{ \includegraphics[width=0.32\textwidth]{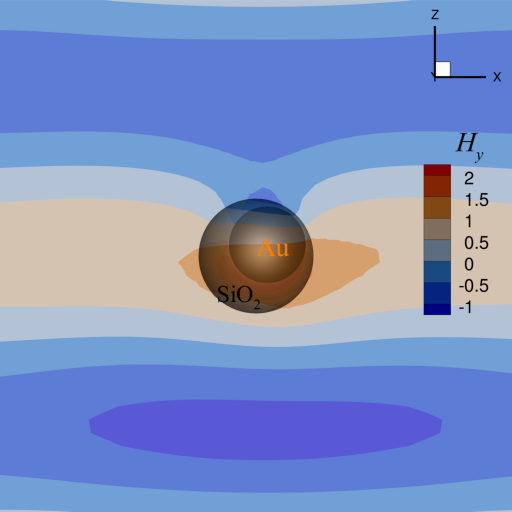}}
\subfloat[$\theta = 315^{\rmo}$]{ \includegraphics[width=0.32\textwidth]{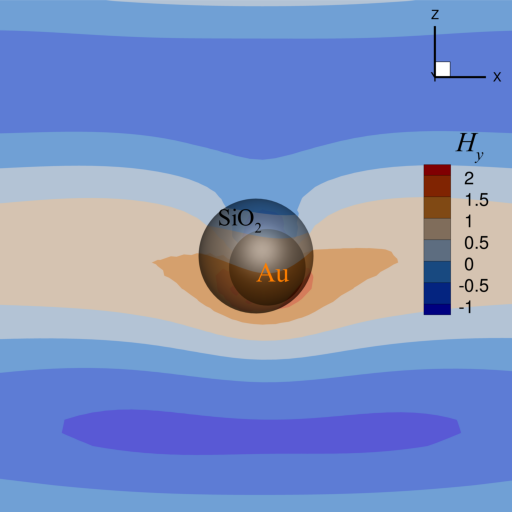}}
\caption{Electric and magnetic fields of when a linearly polarised Gaussian beam is incident on an eccentric Au@SiO$_2$ core-shell particle in air. The wavelength of the beam is 532 nm and its waist radius is 1 $\mu$m. The radius of the Au core is 60 nm, that of the SiO$_2$ shell is 90 nm, and the displacement between the centre of the Au core and the centre of the SiO$_2$ shell, $h=25$ nm. The refractive indices of air, SiO$_2$ and Au are $n_{\text{air}} = 1.0$, $n_{\text{SiO}_{2}} = 1.46$ and $n_{\text{Au}} = 0.54 + \rmi 2.14$, respectively. (a-c) The $x$ component of the electric field; (d-f) The $z$ component of the electric field; (g-i) The $y$ component of the magnetic field at different orientation of the Au core.
}  \label{Fig:EHfield}
\end{figure*}

It is intuitive that the interactions between an eccentric core-shell nanoparticle and external electromagnetic field should be a function of the particle orientation. Take an eccentric Au@SiO$_2$ particle with the shell radius $a_{\text{shell}}=90$ nm and core radius $a_{\text{core}}=60$ nm as an example, when it is under the illumination of a linearly polarised Gaussian beam in air, by using the computational model detailed in Sec. 2 in the main text, Fig.~\ref{Fig:EHfield} shows the total electric and magnetic fields in the surrounding medium and the transmitted electric and magnetic fields in the particles as the centre of Au core is located at different position relative to the centre of the SiO$_2$ shell, when the shell centre is fixed at the focus of the beam. The wavelength of the beam is 532 nm, and the displacement between the centre of the Au core and the centre of the SiO$_2$ shell, $h=25$ nm. The contour plots of the $x$ component of the electric field in the top row of Fig.~\ref{Fig:EHfield} indicate that we will expect a force perpendicular to the wave propagation. Also, under a propagating wave, if the particle size is not very small when compared to the light wavelength, the effects of phase across the particle cannot be ignored. Together with the asymmetry given by the eccentric core-shell allocation, a torque acting on this particle will appear as shown by the contour plots of the $z$ component of the electric field and the $y$ component of the magnetic field in the middle and bottom rows of Fig.~\ref{Fig:EHfield}.

The total electric and magnetic fields in the surrounding medium are the superposition of the incident field and the scattered field as $\boldsymbol{E} = \boldsymbol{E}^{\rminc} + \boldsymbol{E}^{\rmsc}$ and $\boldsymbol{H} = \boldsymbol{H}^{\rminc} + \boldsymbol{H}^{\rmsc}$. Introducing the above relationships into Eqs. (4) and (5) in the main text, we notice that the optical force and torque have three parts: one from the indecent field, one from the scattered field and one from the interactions between the scattered and incident fields as
\begin{subequations}
    \begin{align}
    F^{\rminc}_{i} =& \int_{S_{\text{shell}}} \frac{1}{2} \Big\{ \text{Real} \left[ (D^{\rminc}_{i}E^{\rminc,*}_{j} + E^{\rminc}_{i}D^{\rminc,*}_{j} + B^{\rminc}_{i}H^{\rminc,*}_{j} + H^{\rminc}_{i}B^{\rminc,*}_{j}) n_{j} \right. \nonumber \\
    & \qquad \qquad \qquad \quad \left. - (D^{\rminc}_{j}E^{\rminc,*}_{j} + B^{\rminc}_{j}H^{\rminc,*}_{j} )n_{i}  \right] \Big\} \rmd S \label{eq:force_inc} \\
    F^{\rmsc}_{i} =& \int_{S_{\text{shell}}} \frac{1}{2} \Big\{ \text{Real} \left[ (D^{\rmsc}_{i}E^{\rmsc,*}_{j} + E^{\rmsc}_{i}D^{\rmsc,*}_{j} + B^{\rmsc}_{i}H^{\rmsc,*}_{j} + H^{\rmsc}_{i}B^{\rmsc,*}_{j}) n_{j} \right. \nonumber \\
    & \qquad \qquad \qquad \quad \left. - (D^{\rmsc}_{j}E^{\rmsc,*}_{j} + B^{\rmsc}_{j}H^{\rmsc,*}_{j} )n_{i}  \right] \Big\} \rmd S \label{eq:force_sca} \\
    F^{\rmext}_{i} =& \int_{S_{\text{shell}}} \frac{1}{2} \Big\{ \text{Real} \left[ (D^{\rminc}_{i}E^{\rmsc,*}_{j} + E^{\rminc}_{i}D^{\rmsc,*}_{j} + B^{\rminc}_{i}H^{\rmsc,*}_{j} + H^{\rminc}_{i}B^{\rmsc,*}_{j}) n_{j} \right. \nonumber \\
    & \qquad \qquad \qquad \quad + (D^{\rmsc}_{i}E^{\rminc,*}_{j} + E^{\rmsc}_{i}D^{\rminc,*}_{j} + B^{\rmsc}_{i}H^{\rminc,*}_{j} + H^{\rmsc}_{i}B^{\rminc,*}_{j}) n_{j}  \nonumber \\
    & \qquad \qquad \qquad \quad \left. - (D^{\rminc}_{j}E^{\rmsc,*}_{j} + B^{\rminc}_{j}H^{\rmsc,*}_{j} )n_{i} - (D^{\rmsc}_{j}E^{\rminc,*}_{j} + B^{\rmsc}_{j}H^{\rminc,*}_{j} )n_{i}  \right] \Big\} \rmd S \label{eq:force_ext}
    \end{align}
\end{subequations}\label{eq:force}
\begin{subequations}\label{eq:torque}
    \begin{align}
    N^{\rminc}_{i} =& \int_{S_{\text{shell}}} \varepsilon_{ijk} r^{c}_{j}\frac{1}{2} \Big\{ \text{Real} \left[ (D^{\rminc}_{k}E^{\rminc,*}_{l} + E^{\rminc}_{k}D^{\rminc,*}_{l} + B^{\rminc}_{k}H^{\rminc,*}_{l} + H^{\rminc}_{k}B^{\rminc,*}_{l}) n_{l} \right. \nonumber \\
    & \qquad \qquad \qquad \quad \left. - (D^{\rminc}_{l}E^{\rminc,*}_{l} + B^{\rminc}_{l}H^{\rminc,*}_{l} )n_{k}  \right]  \Big\} \rmd S \label{eq:torque_inc} \\
    N^{\rmsc}_{i} =& \int_{S_{\text{shell}}} \varepsilon_{ijk} r^{c}_{j}\frac{1}{2}  \Big\{ \text{Real} \left[ (D^{\rmsc}_{k}E^{\rmsc,*}_{l} + E^{\rmsc}_{k}D^{\rmsc,*}_{l} + B^{\rmsc}_{k}H^{\rmsc,*}_{l} + H^{\rmsc}_{k}B^{\rmsc,*}_{l}) n_{l} \right. \nonumber \\
    & \qquad \qquad \qquad \quad \left. - (D^{\rmsc}_{l}E^{\rmsc,*}_{l} + B^{\rmsc}_{l}H^{\rmsc,*}_{l} )n_{k}  \right]  \Big\} \rmd S \label{eq:torque_sca} \\
    N^{\rmext}_{i} =& \int_{S_{\text{shell}}} \varepsilon_{ijk} r^{c}_{j}\frac{1}{2}  \Big\{ \text{Real} \left[ (D^{\rminc}_{k}E^{\rmsc,*}_{l} + E^{\rminc}_{k}D^{\rmsc,*}_{l} + B^{\rminc}_{k}H^{\rmsc,*}_{l} + H^{\rminc}_{k}B^{\rmsc,*}_{l}) n_{l} \right. \nonumber \\
    & \qquad \qquad \qquad \quad + (D^{\rmsc}_{k}E^{\rminc,*}_{l} + E^{\rmsc}_{k}D^{\rminc,*}_{l} + B^{\rmsc}_{k}H^{\rminc,*}_{l} + H^{\rmsc}_{k}B^{\rminc,*}_{l}) n_{l}  \nonumber \\
    & \qquad \qquad \qquad \quad \left. - (D^{\rminc}_{l}E^{\rmsc,*}_{l} + B^{\rminc}_{l}H^{\rmsc,*}_{l} )n_{k} - (D^{\rmsc}_{l}E^{\rminc,*}_{l} + B^{\rmsc}_{l}H^{\rminc,*}_{l} )n_{k}  \right]  \Big\} \rmd S \label{eq:torque_ext}
    \end{align}
\end{subequations}

Based on the numerical experiments set up in our work, since the shell is spherical and its centre is fixed at the focus of the incident Gaussian beam, the force and torque due to the incident fields $\boldsymbol{E}^{\rminc}$ and $\boldsymbol{H}^{\rminc}$ are zeros: $\boldsymbol{F}^{\rminc} = \boldsymbol{0}$ and $\boldsymbol{N}^{\rminc} = \boldsymbol{0}$. It is worth mentioning that if the particle is not located at the focus of beam, there will be optical force generated from the incident field (gradient force). As such, in this work, we investigated the total optical force and torque, $\boldsymbol{F}$ and $\boldsymbol{N}$ defined in Eqs.~(4) and (5) in the main text, respectively, and those from the scattered field, $\boldsymbol{F}^{\rmsc}$ and $\boldsymbol{N}^{\rmsc}$ defined in Eqs.~(\ref{eq:force_sca}) and (\ref{eq:torque_sca}), respectively, and those from the interaction between the scattered field and the incident field $\boldsymbol{F}^{\rmext}$ and $\boldsymbol{N}^{\rmext}$ defined in Eqs.~(\ref{eq:force_ext}) and (\ref{eq:torque_ext}), respectively.

\subsection{Optical force along the beam propagation direction}

\begin{figure*}[!t]
\centering{}
\subfloat[]{ \includegraphics[width=0.32\textwidth]{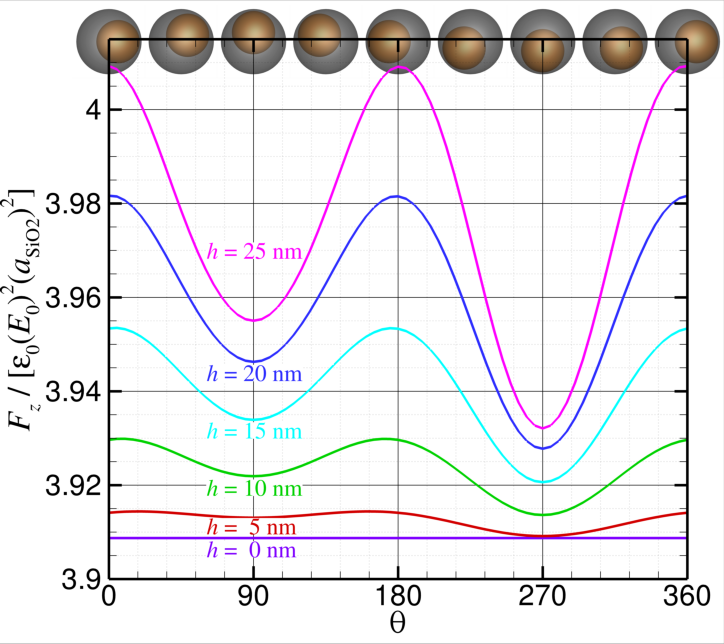}}
\subfloat[]{ \includegraphics[width=0.32\textwidth]{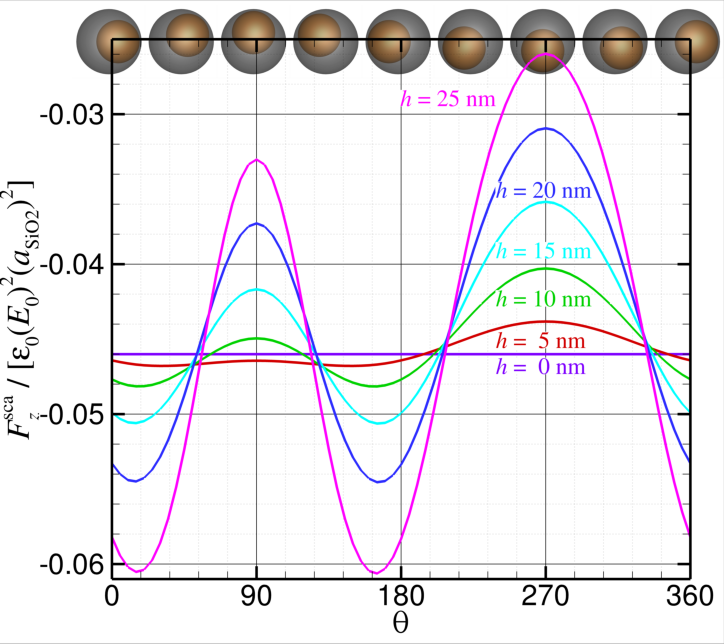}}
\subfloat[]{ \includegraphics[width=0.32\textwidth]{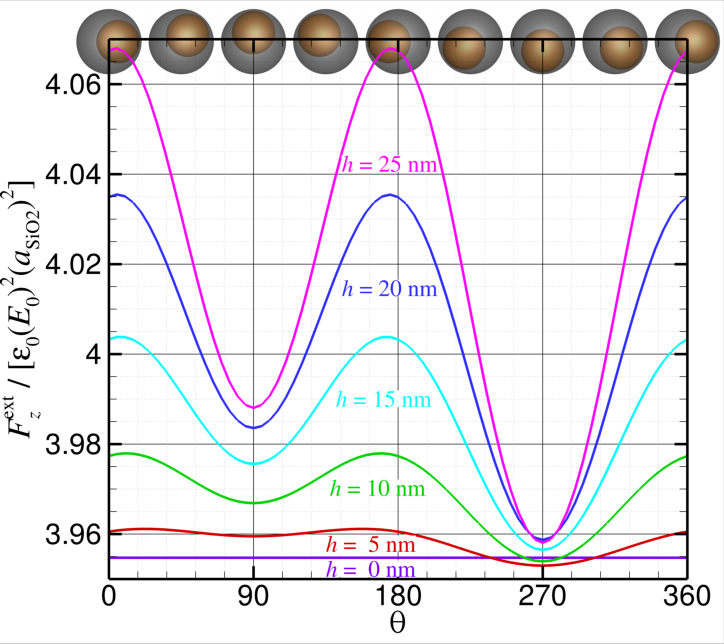}} \\
\subfloat[]{ \includegraphics[width=0.32\textwidth]{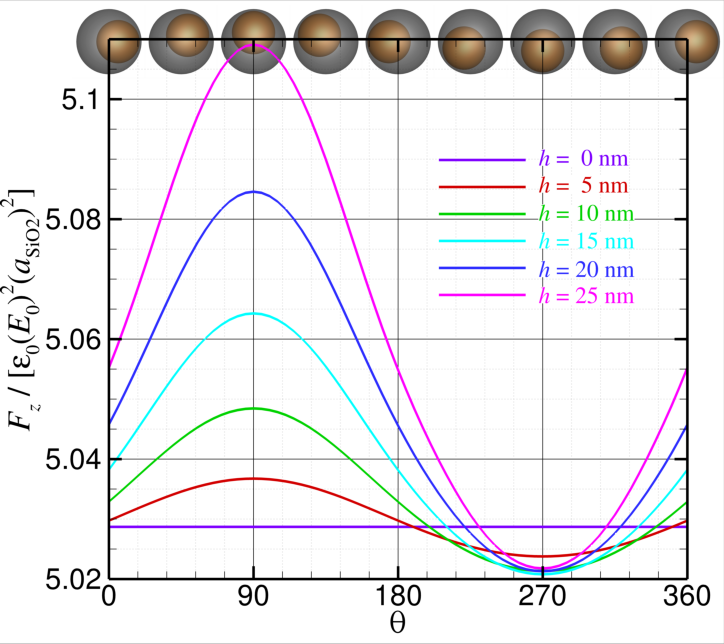}}
\subfloat[]{ \includegraphics[width=0.32\textwidth]{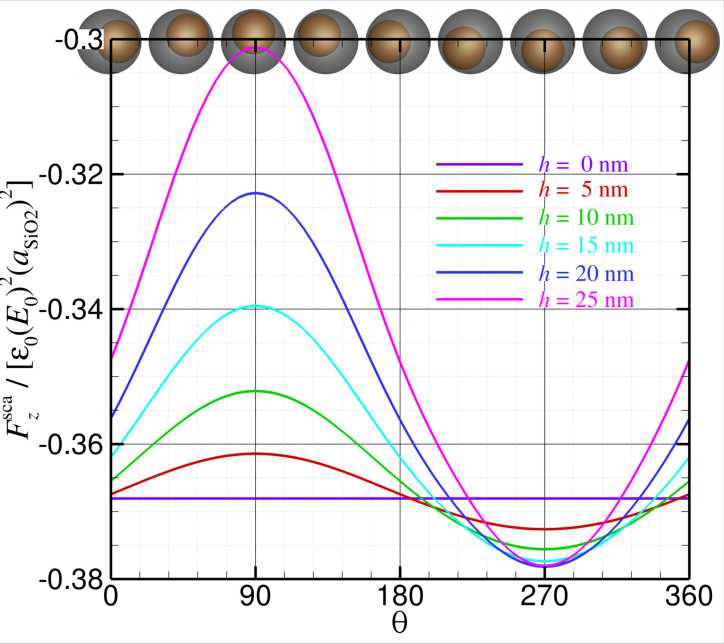}}
\subfloat[]{ \includegraphics[width=0.32\textwidth]{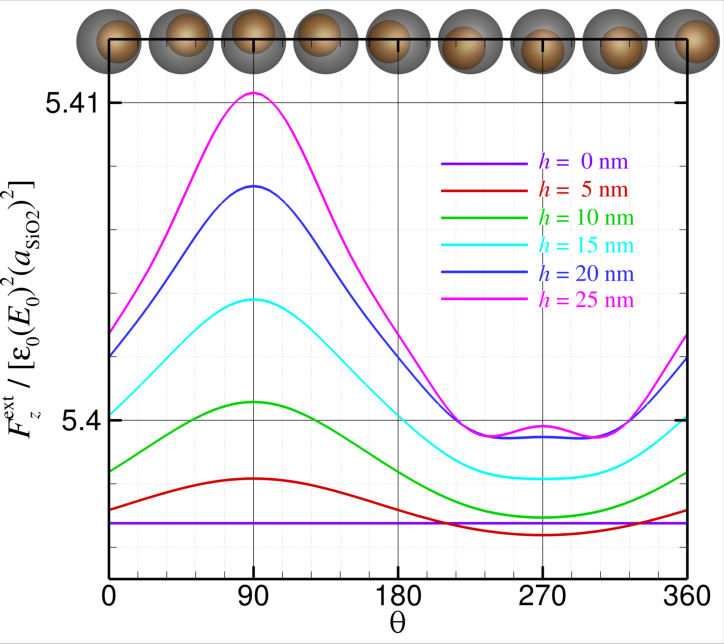}}
\caption{Optical force along the direction of incident beam propagation on the eccentric Au@SiO$_2$ core-shell particle under the linearly polarised Gaussian beam illumination. (a-c) in air with $n_{\text{air}}=1$; (d-f) in water with $n_{\text{water}}=1.33$. The shell radius is $a_{\text{shell}}=90$ nm, the core radius is $a_{\text{core}}=60$ nm, the beam wavelength is $\lambda=532$ nm, and the beam waist radius is $w_0=1$ $\mu$m.
}  \label{Fig:Fz_h}
\end{figure*}

Let us first consider the optical force along the beam propagation or the trapping force. Taking an Au@SiO$_2$ eccentric core-shell particle with the shell radius $a_{\text{shell}}=90$ nm and core radius $a_{\text{core}}=60$ nm which shell centre is at the focal point of the Gaussian beam as an example. Fig.~\ref{Fig:Fz_h} shows how the asymmetry, $h$, can affect the trapping force $F_z$ and its components $F_z^{\rmsc}$ and $F_z^{\rmext}$ when an eccentric Au@SiO$_{2}$ core-shell particle illuminated by a beam with wavelength $\lambda=532$ nm. The top row of Fig.~\ref{Fig:Fz_h} shows the variations of $F_z$ , $F_z^{\rmsc}$ and $F_z^{\rmext}$ along with the Au core orientation, $\theta$ when the surrounding medium is air with $n_{\text{air}}=1$. In Fig.~\ref{Fig:Fz_h}(a), we can see that when the distance, $h$, between the Au core-centre and the SiO$_2$ shell-centre increases, the total optical force along the wave propagation, $F_z$ becomes larger and larger at all orientation angle $\theta$. It is noticeable that the force curves have two local minima when the orientation angle $\theta=90^{\rmo}$ and $\theta=270^{\rmo}$, respectively, where $F_{z}$ is larger at $\theta=90^{\rmo}$ compared to that at $\theta=270^{\rmo}$. The trends of the optical force along the wave propagation due to the interaction between the scattered field and the incident field, $F_z^{\rmext}$ behaves the same, as shown in Fig.~\ref{Fig:Fz_h}(c). It is noticeable that the magnitude of this part of the optical force, $F_z^{\rmext}$ is higher than the net optical force, $F_z$. This is due to the fact that the optical force from the scattered field shows a tractor effect (opposite to the wave propagation direction), as shown in Figs~\ref{Fig:Fz_h}(b). Nevertheless, the optical force from the scattered field, $F_z^{\rmsc}$ is much smaller relative to that from the interaction between the incident and the scattered field, $F_z^{\rmext}$.

\begin{figure*}[!t]
\centering{}
\subfloat[Au@SiO$_{2}$]{ \includegraphics[width=0.32\textwidth]{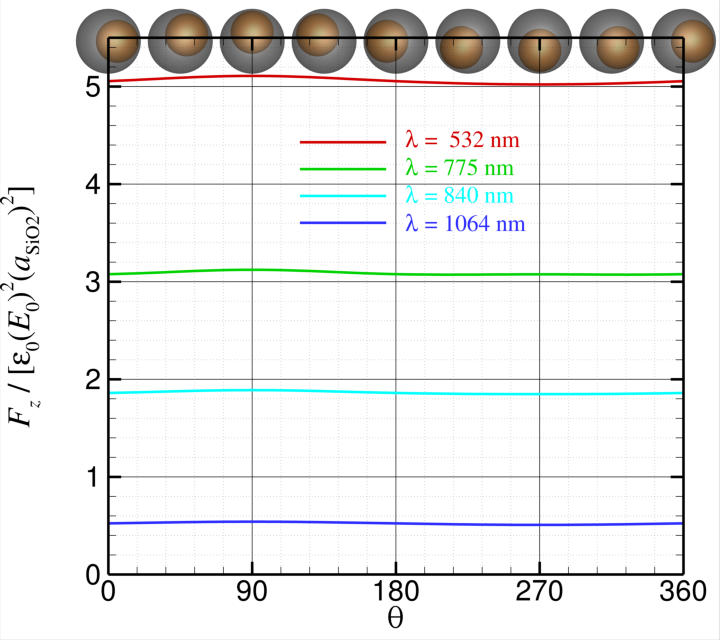}}
\subfloat[Au@SiO$_{2}$]{ \includegraphics[width=0.32\textwidth]{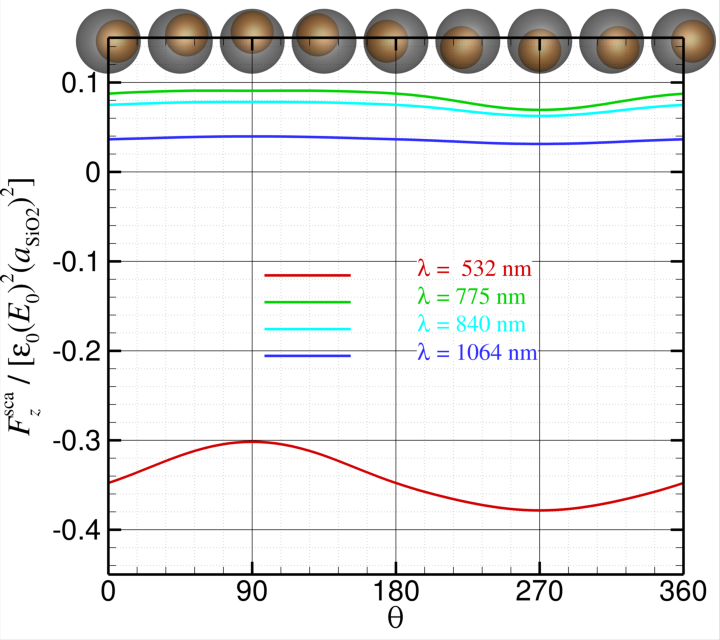}}
\subfloat[Au@SiO$_{2}$]{ \includegraphics[width=0.32\textwidth]{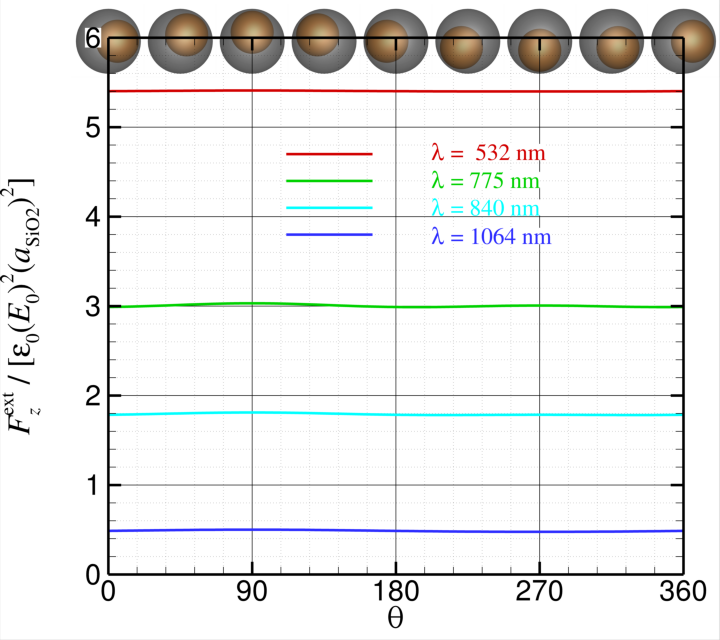}} \\
\subfloat[TiO$_2$@SiO$_{2}$]{ \includegraphics[width=0.32\textwidth]{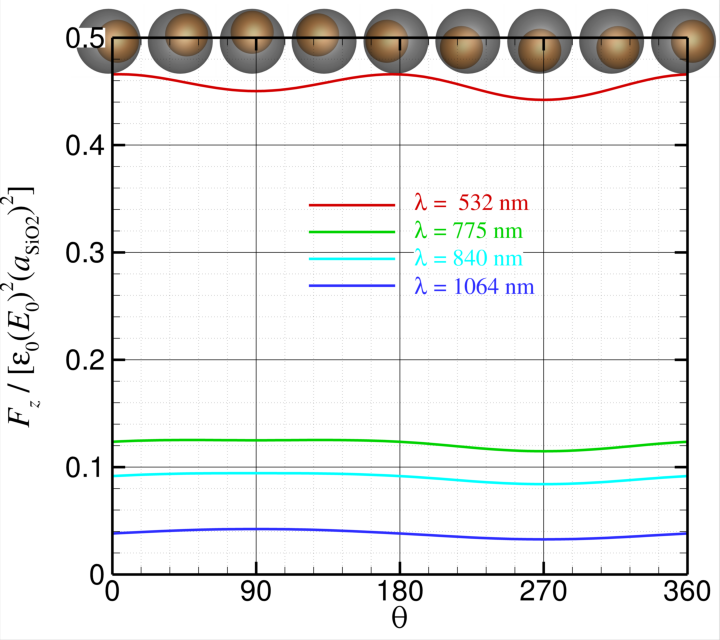}}
\subfloat[TiO$_2$@SiO$_{2}$]{ \includegraphics[width=0.32\textwidth]{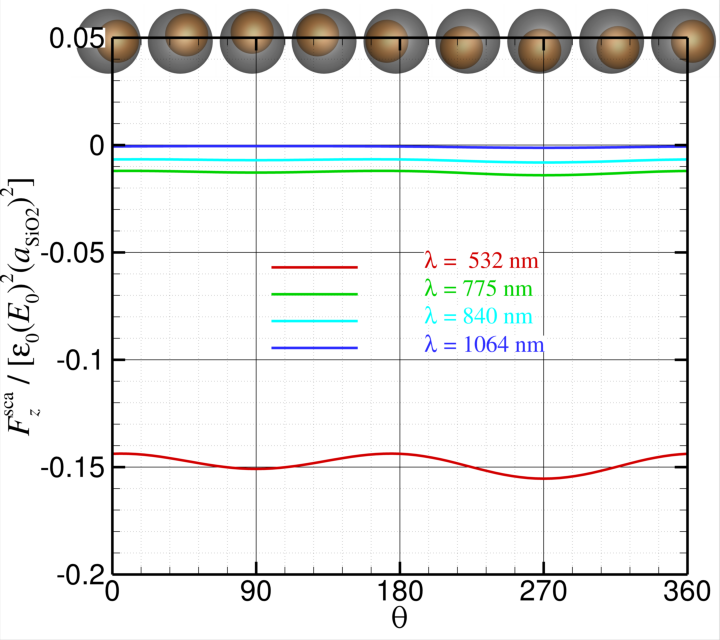}}
\subfloat[TiO$_2$@SiO$_{2}$]{ \includegraphics[width=0.32\textwidth]{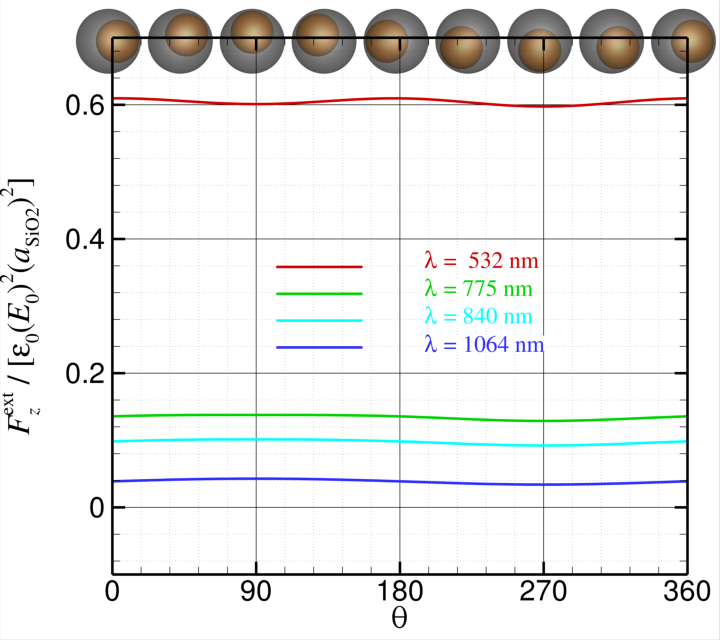}} \\
\subfloat[SiO$_2$@TiO$_{2}$]{ \includegraphics[width=0.32\textwidth]{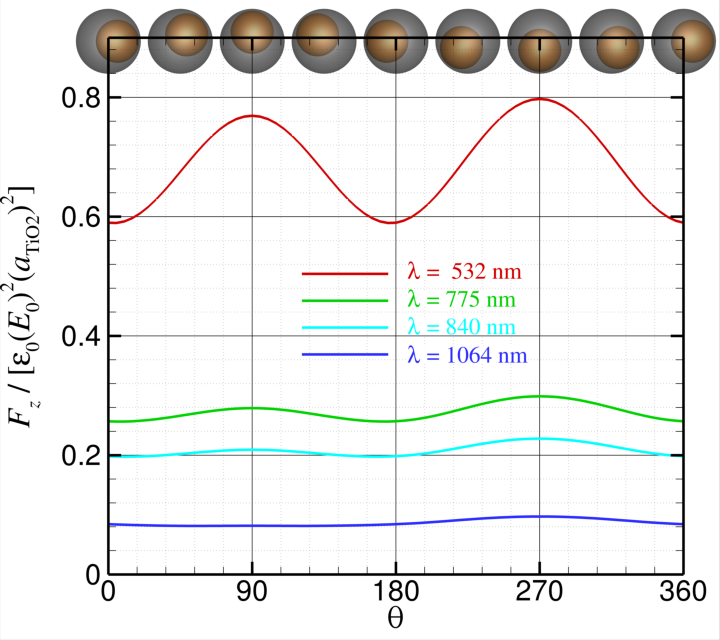}}
\subfloat[SiO$_2$@TiO$_{2}$]{ \includegraphics[width=0.32\textwidth]{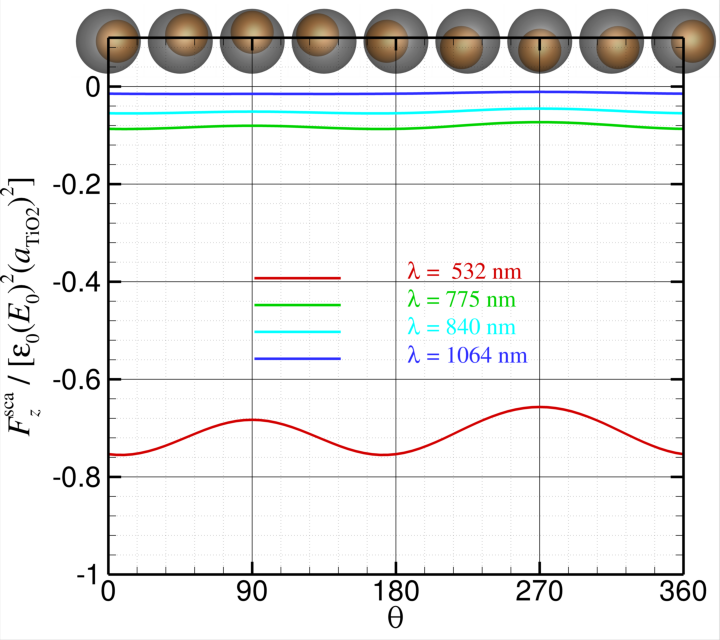}}
\subfloat[SiO$_2$@TiO$_{2}$]{ \includegraphics[width=0.32\textwidth]{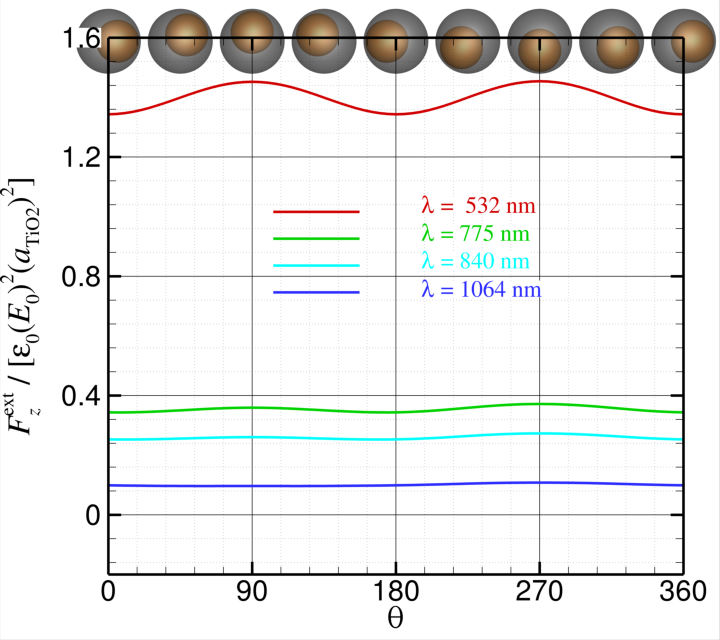}}
\caption{Optical force along the direction of incident beam propagation on three types of eccentric core-shell particles in water under a linearly polarised Gaussian beam illumination with beam waist radius as $w_0=1$ $\mu$m. The geometrical features of the core-shell particle are $a_{\text{shell}}=90$ nm, $a_{\text{core}}=60$ nm and $h=25$ nm.
}  \label{Fig:Fz_lambda}
\end{figure*}

When the surrounding medium is water with $n_{\text{water}} = 1.33$, the variations of the optical forces, $F_{z}$ due to the asymmetry of the Au core change accordingly because of the difference of the ratio of the relative refractive indices between the SiO$_2$ shell and the surrounding medium. One obvious difference is that $\theta=90^{\rmo}$ becomes the orientation of Au core corresponding to the maximum optical force along the wave propagation, as shown in Fig.~\ref{Fig:Fz_h}(d). Also, when the asymmetry $h$ of the eccentric particle becomes larger, the net optical force along the beam propagation, $F_z$ increases at most orientations of the Au core except for a small range around $\theta = 270^{\rmo}$ which is the effect from the force component due to the scattered field, $F_z^{\rmsc}$ as displayed in Fig.~\ref{Fig:Fz_h}(e). When comparing Figs.~\ref{Fig:Fz_h}(e-f), we can see that the optical force from the interaction between the incident and the scattered field, $F_z^{\rmext}$ dominates that from the scattered field, $F_z^{\rmsc}$. Converting the magnitude of the optical trapping force, $F_z$ in Fig.~\ref{Fig:Fz_h}(d), we obtain the $Q$-factor, $Q = F_z \, c/(n_{\text{medium}}P_0)$ with $c$ being the speed of light, larger than 0.029 which is reasonable. 

Fig.~\ref{Fig:Fz_lambda} presents the effect of wavelength on the optical force along the Gaussian beam propagation for three types of eccentric core-shell particles in water: Au@SiO$_{2}$, TiO$_2$@SiO$_{2}$ and SiO$_2$@TiO$_{2}$. Four wavelengths, $\lambda=532$ nm, $\lambda=775$ nm, $\lambda=840$ nm and $\lambda=1064$ nm are under consideration when the linearly polarised Gaussian beam waist radius is fixed as $w_0=1$ $\mu$m. As the wavelength becomes larger and larger, the magnitudes of the net optical force, $F_z$ decreases. From the first and second rows of Fig.~\ref{Fig:Fz_lambda}, we can clearly see that the optical force from the interaction between the scattered and incident fields, $F_z^{\rmext}$ dominates the force from the scattered field, $F_z^{\rmsc}$ for Au@SiO$_{2}$ and TiO$_2$@SiO$_{2}$. However, for SiO$_2$@TiO$_{2}$, the magnitude of these two parts are in the same order and they are competing with each other. The net force $F_z$ shown in Fig.~\ref{Fig:Fz_lambda}(g) indicates that the optical force from the interaction between the scattered and incident fields, $F_z^{\rmext}$ overcomes the force from the scattered field, $F_z^{\rmsc}$.

\subsection{Optical force perpendicular to the beam propagation}

\begin{figure*}[t]
\centering{}
\subfloat[]{ \includegraphics[width=0.32\textwidth]{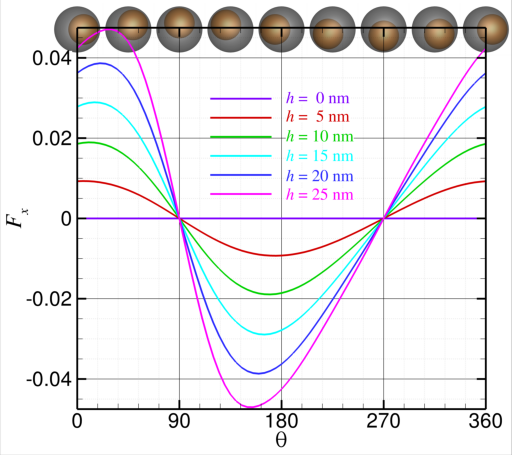}}
\subfloat[]{ \includegraphics[width=0.32\textwidth]{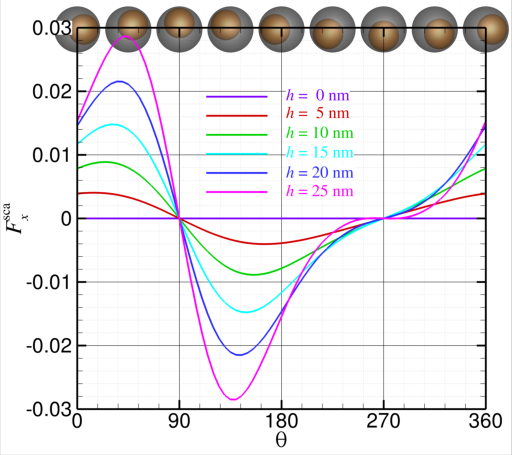}}
\subfloat[]{ \includegraphics[width=0.32\textwidth]{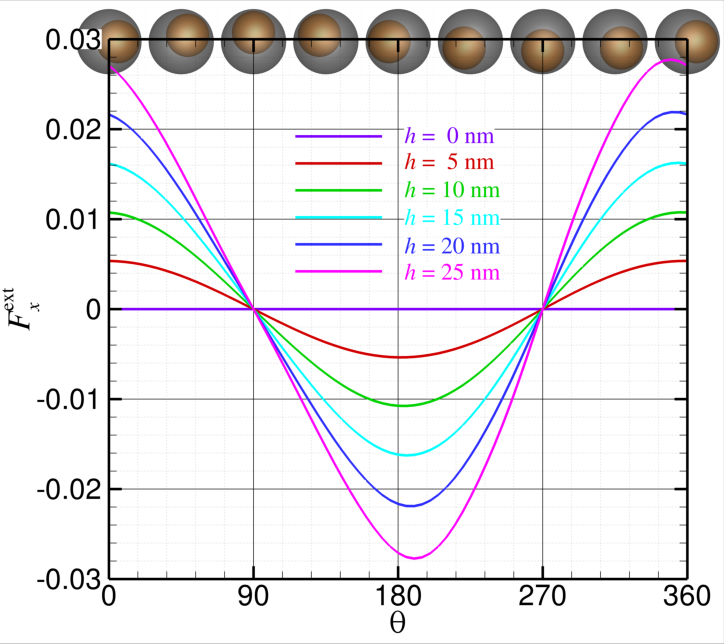}} \\
\subfloat[]{ \includegraphics[width=0.32\textwidth]{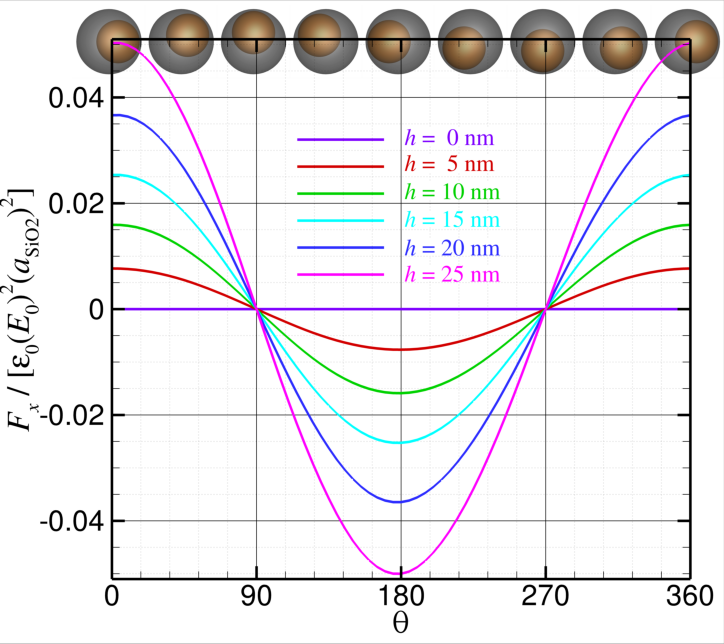}}
\subfloat[]{ \includegraphics[width=0.32\textwidth]{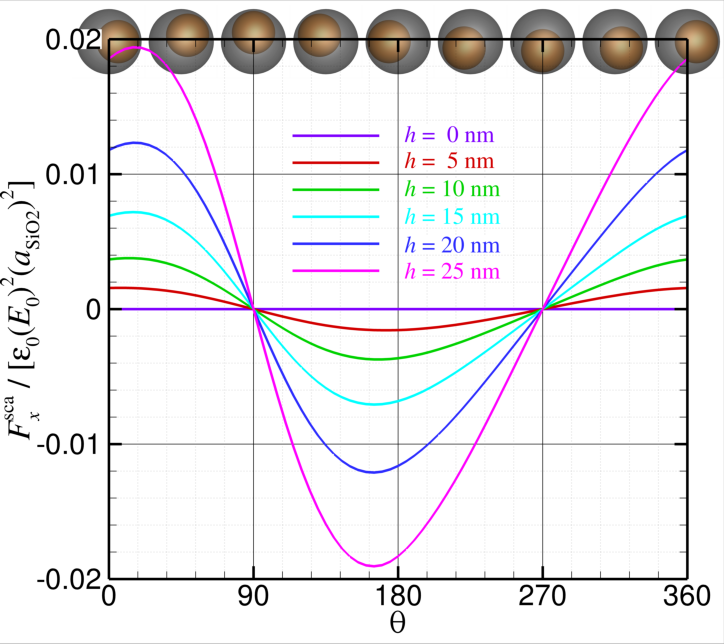}}
\subfloat[]{ \includegraphics[width=0.32\textwidth]{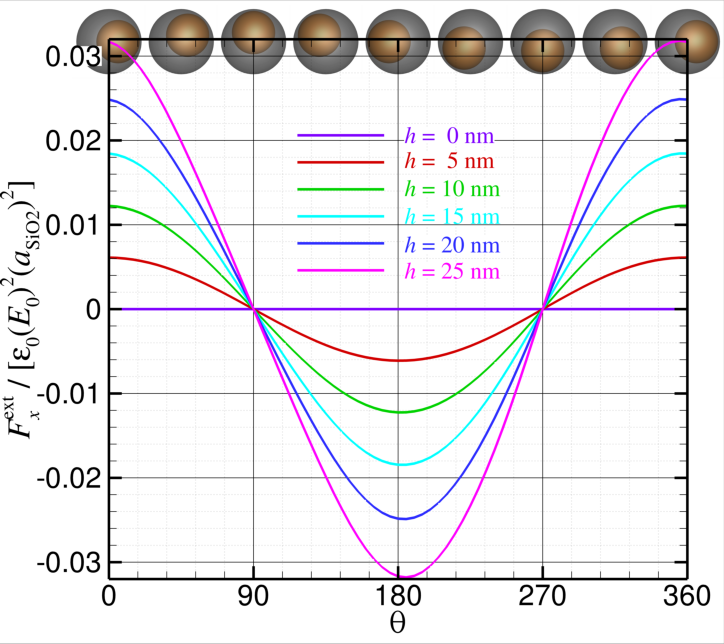}}
\caption{Optical force perpendicular to the direction of incident beam propagation on the eccentric Au@SiO$_2$ core-shell particle under the linearly polarised Gaussian beam illumination. (a-c) in air with $n_{\text{air}}=1$; (d-f) in water with $n_{\text{water}}=1.33$. The shell radius is $a_{\text{shell}}=90$ nm, the core radius is $a_{\text{core}}=60$ nm, the beam wavelength is $\lambda=532$ nm, and the beam waist radius is $w_0=1$ $\mu$m.
}  \label{Fig:Fx_h}
\end{figure*}
\begin{figure*}[!ht]
\centering{}
\subfloat[Au@SiO$_{2}$]{ \includegraphics[width=0.32\textwidth]{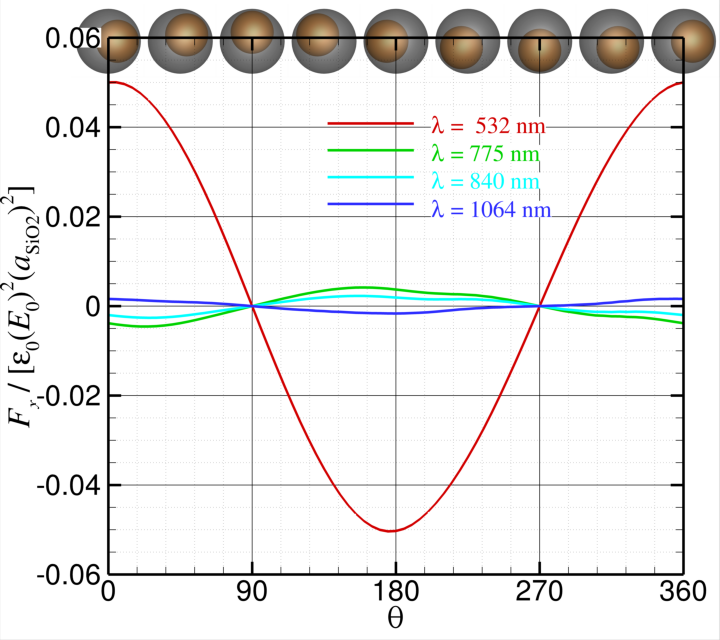}}
\subfloat[Au@SiO$_{2}$]{ \includegraphics[width=0.32\textwidth]{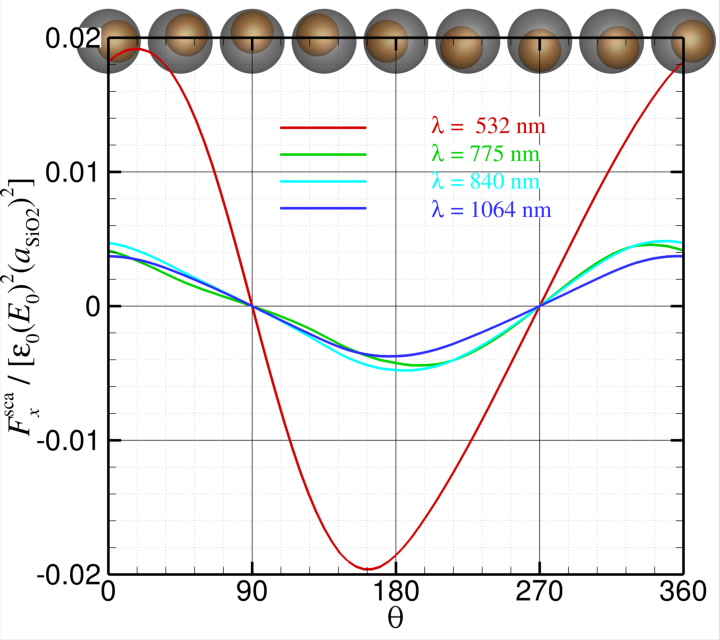}}
\subfloat[Au@SiO$_{2}$]{ \includegraphics[width=0.32\textwidth]{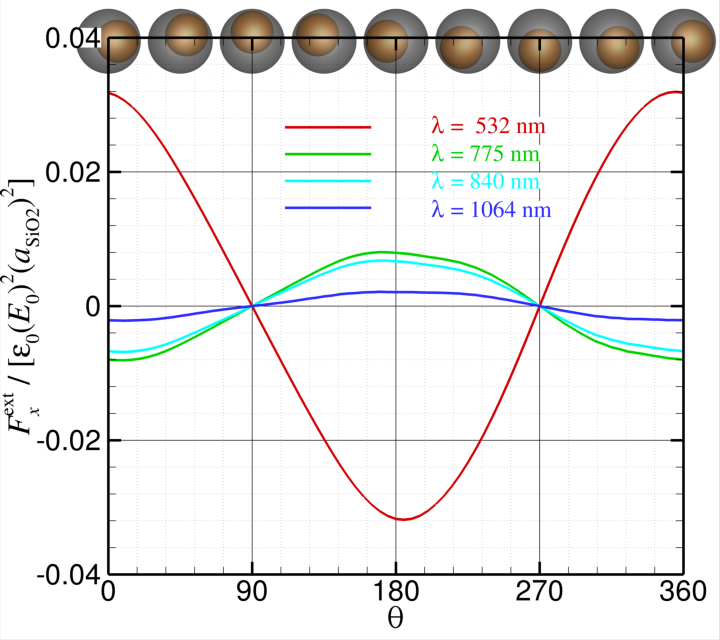}} \\
\subfloat[TiO$_2$@SiO$_{2}$]{ \includegraphics[width=0.32\textwidth]{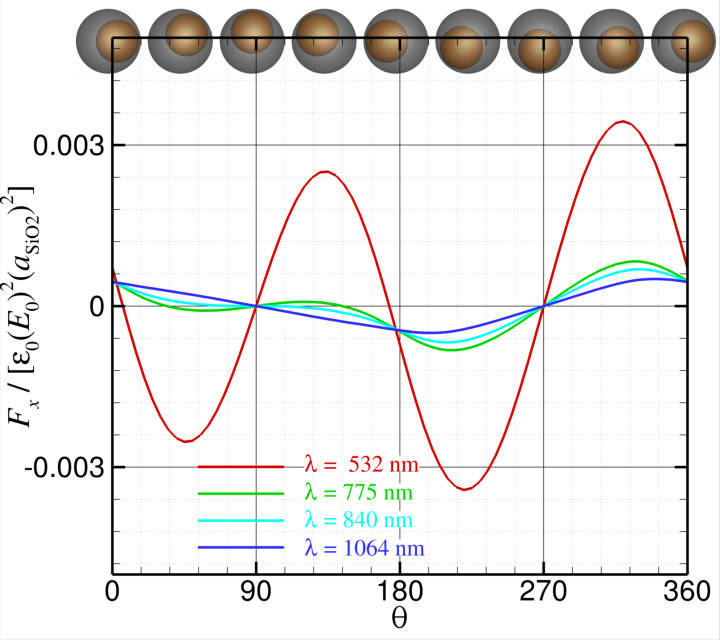}}
\subfloat[TiO$_2$@SiO$_{2}$]{ \includegraphics[width=0.32\textwidth]{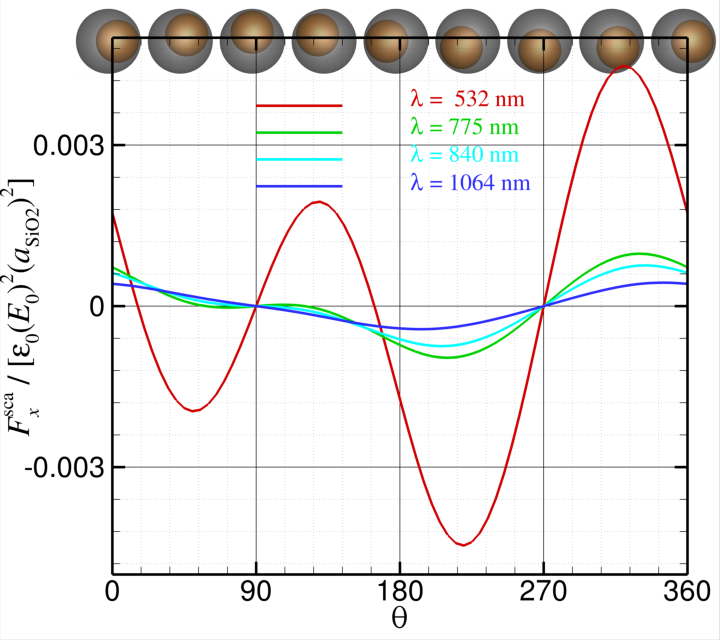}}
\subfloat[TiO$_2$@SiO$_{2}$]{ \includegraphics[width=0.32\textwidth]{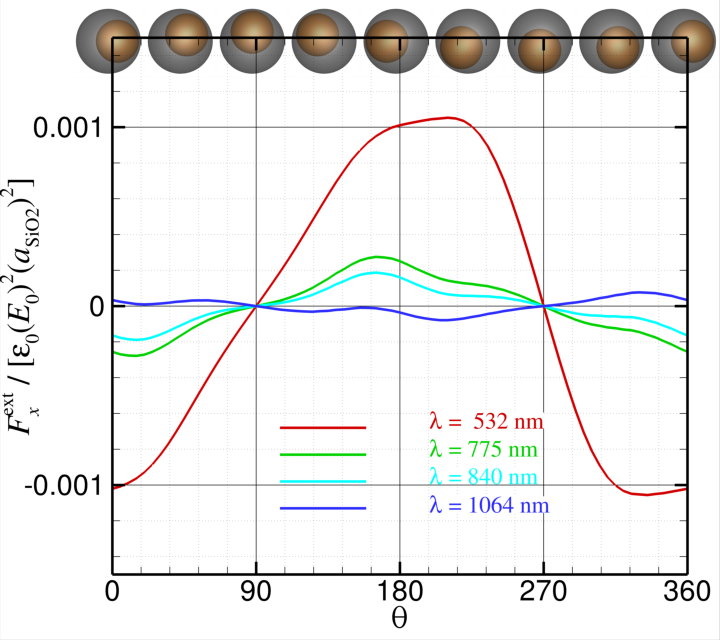}} \\
\subfloat[SiO$_2$@TiO$_{2}$]{ \includegraphics[width=0.32\textwidth]{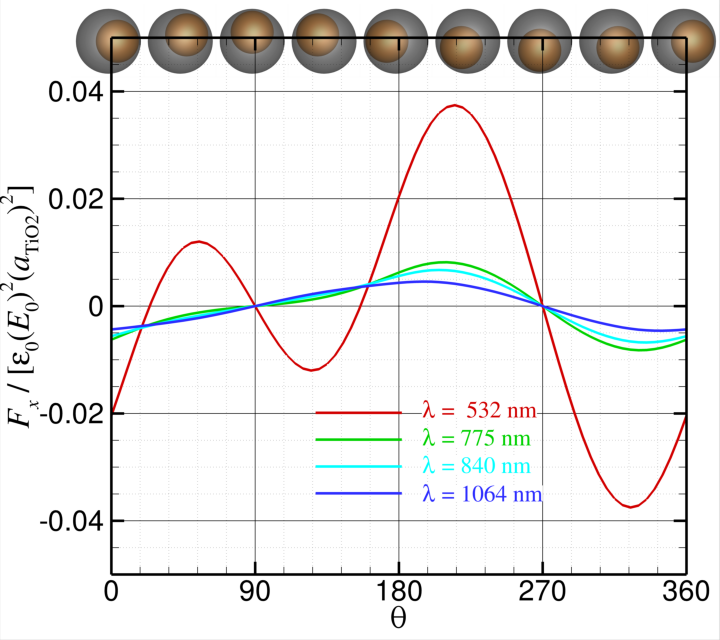}}
\subfloat[SiO$_2$@TiO$_{2}$]{ \includegraphics[width=0.32\textwidth]{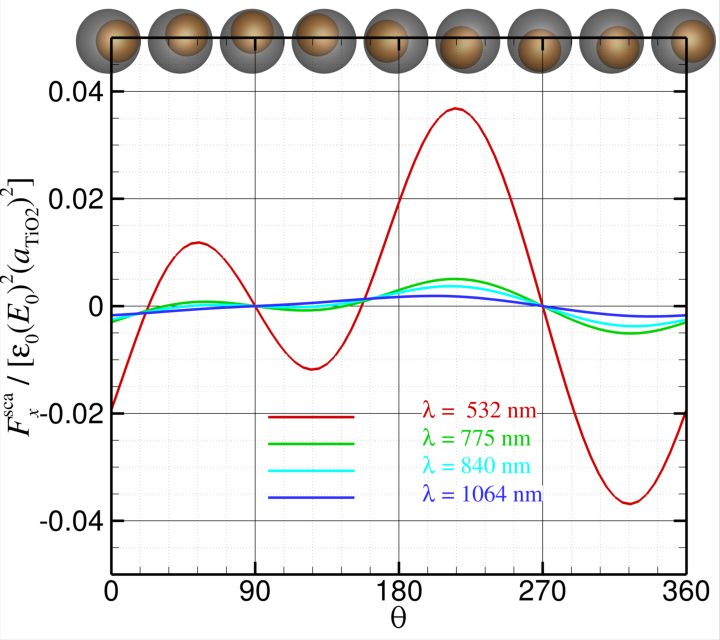}}
\subfloat[SiO$_2$@TiO$_{2}$]{ \includegraphics[width=0.32\textwidth]{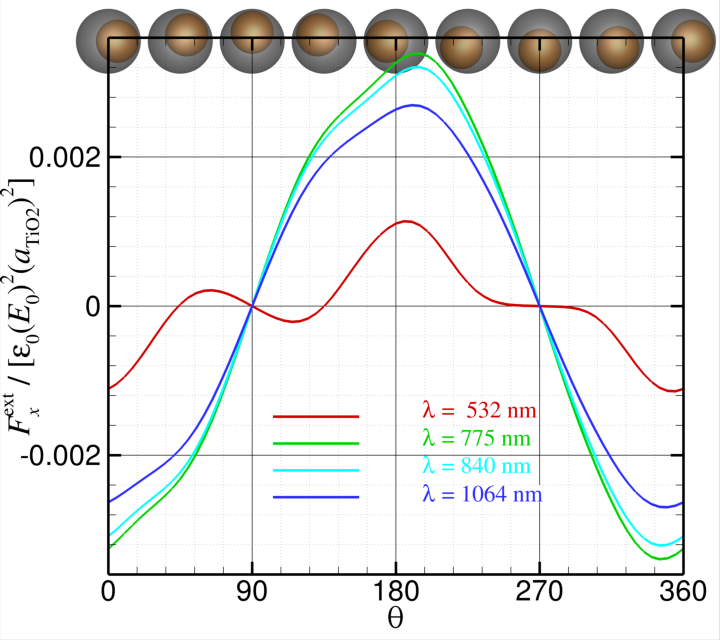}}
\caption{Optical force perpendicular to the direction of incident beam propagation on three types of eccentric core-shell particles in water under a linearly polarised Gaussian beam illumination with beam waist radius as $w_0=1$ $\mu$m. The geometrical features of the core-shell particle are $a_{\text{shell}}=90$ nm, $a_{\text{core}}=60$ nm and $h=25$ nm.
}  \label{Fig:Fx_lambda}
\end{figure*}

As the core particle is not concentric with the centre of the shell, there will be an optical force perpendicular to the beam propagation direction. Fig.~\ref{Fig:Fx_h} displays the optical force perpendicular to a linearly polarised Gaussian beam with its light wavelength as $\lambda=532$ nm illuminating on an eccentric Au@SiO$_2$ core-shell particle. We, firstly, would like to point out that compared to the optical force along the beam propagation axis $F_z$, $F_x$ can be deemed as a secondary effect since the magnitude of $F_x$ is 2 orders lower than that of $F_z$.

The variation trends of such a force, $F_x$ and its components, $F_x^{\rmsc}$ and $F_x^{\rmext}$ with respect to the orientation of the Au core, $\theta$ are quite similar for air or water as the surrounding medium. It is intuitive to consider that the maximum value of this optical force, $F_x$ appears when the core particle is at the maximum displacement from the beam axis at $\theta=0^{\rmo}$ and $\theta=180^{\rmo}$ which is demonstrated in Fig.~\ref{Fig:Fx_h}. If the surrounding medium is air, the orientation of the Au core corresponding to the largest value of $F_x$ is shifted a little away from $\theta=0^{\rmo}$ or $\theta=180^{\rmo}$ as displayed in ~\ref{Fig:Fx_h}(a). This is because the change of the refractive index of the surrounding medium tuned the scattered field that consequently affects the optical force, $F_x$. Unlike the force along the beam propagation, $F_z$ in which the contribution of the interaction between the scattered and incident fields is dominating, for the optical force perpendicular to the beam direction, $F_x$, both the force from the scattered field, $F_x^{\rmsc}$ and that from the interaction of the scattered and incident fields, $F_x^{\rmext}$ have similar contributions to the net $F_x$ for the Au@SiO$_2$ core-shell particle when $\lambda=532$ nm.

The optical force perpendicular to beam propagation under different wavelengths on three types of eccentric core-shell particles with $h=25$ nm are shown in Fig.~{\ref{Fig:Fx_lambda}}. For the Au@SiO$_2$ core-shell particle at long wavelengths, the force from the scattered field, $F_x^{\rmsc}$ competes with that from the interaction of the scattered and incident fields, $F_x^{\rmext}$, as shown in Figs.~{\ref{Fig:Fx_lambda}}(b-c). As to the TiO$_2$@SiO$_{2}$ and SiO$_2$@TiO$_{2}$ core-shell particles, the net force, $F_x$ is dominated by the contribution from the scattered field, $F_x^{\rmsc}$ for all the wavelengths under consideration, as displayed in the second and third rows of Fig.~{\ref{Fig:Fx_lambda}}.

\begin{figure*}[t]
\centering{}
\subfloat[]{ \includegraphics[width=0.32\textwidth]{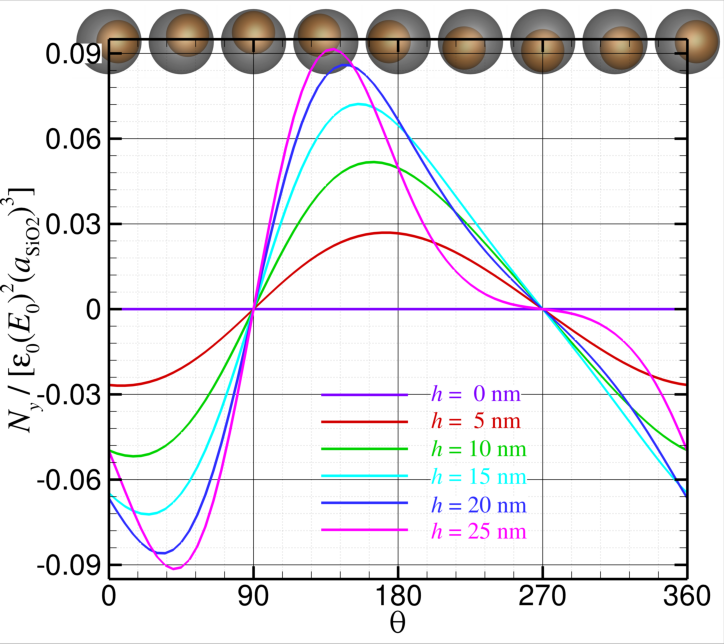}}
\subfloat[]{ \includegraphics[width=0.32\textwidth]{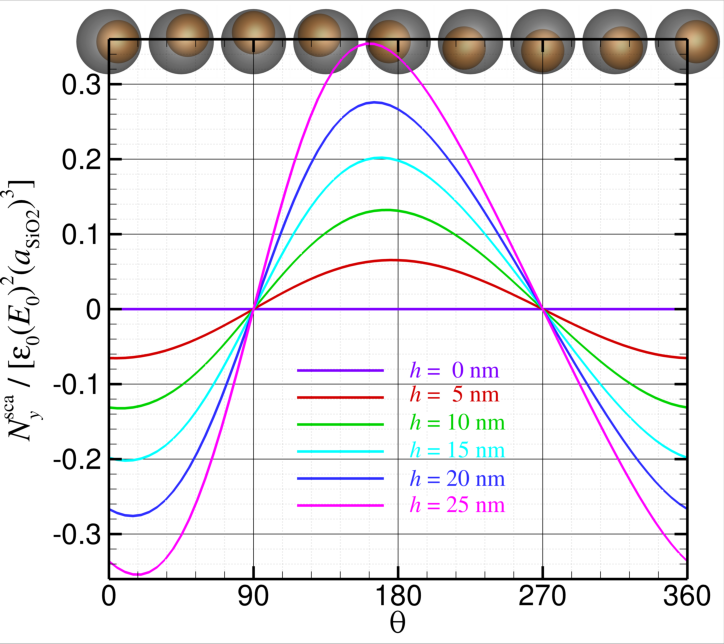}}
\subfloat[]{ \includegraphics[width=0.32\textwidth]{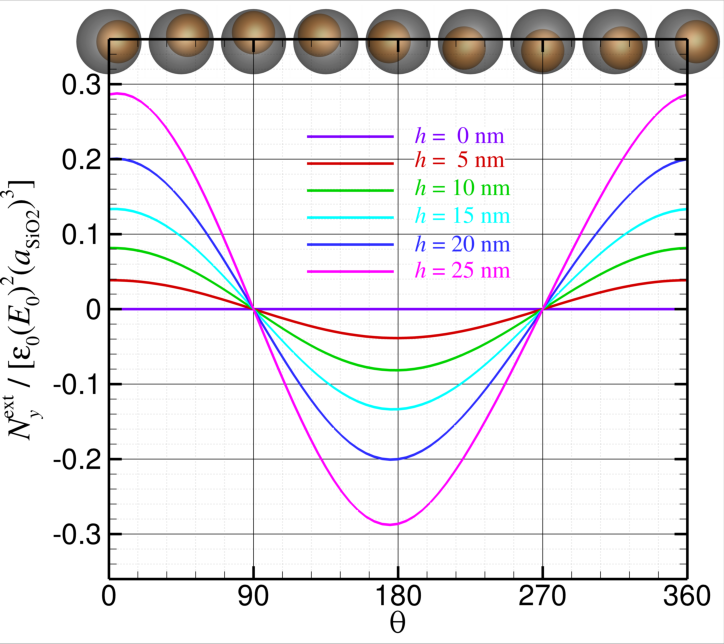}} \\
\subfloat[]{ \includegraphics[width=0.32\textwidth]{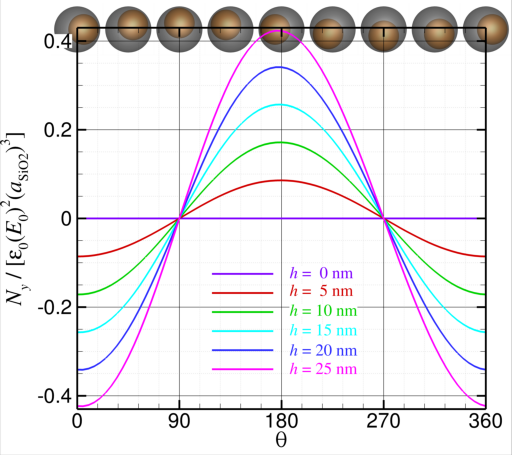}}
\subfloat[]{ \includegraphics[width=0.32\textwidth]{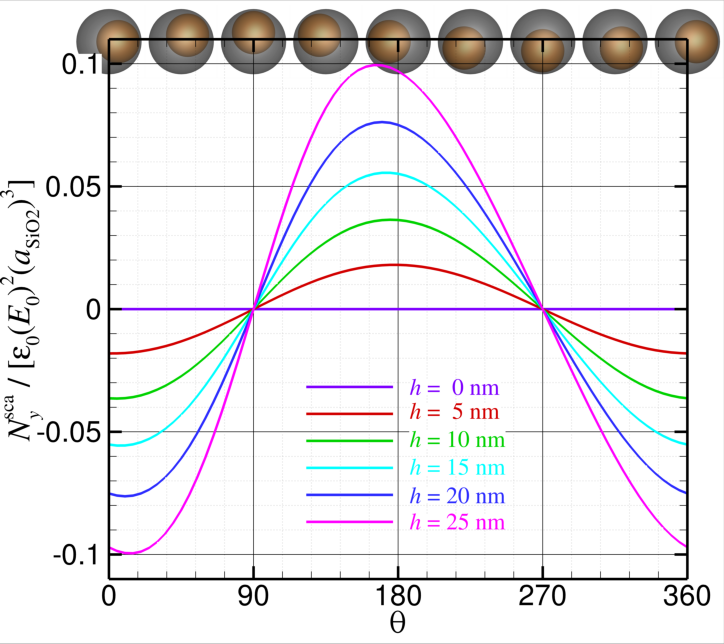}}
\subfloat[]{ \includegraphics[width=0.32\textwidth]{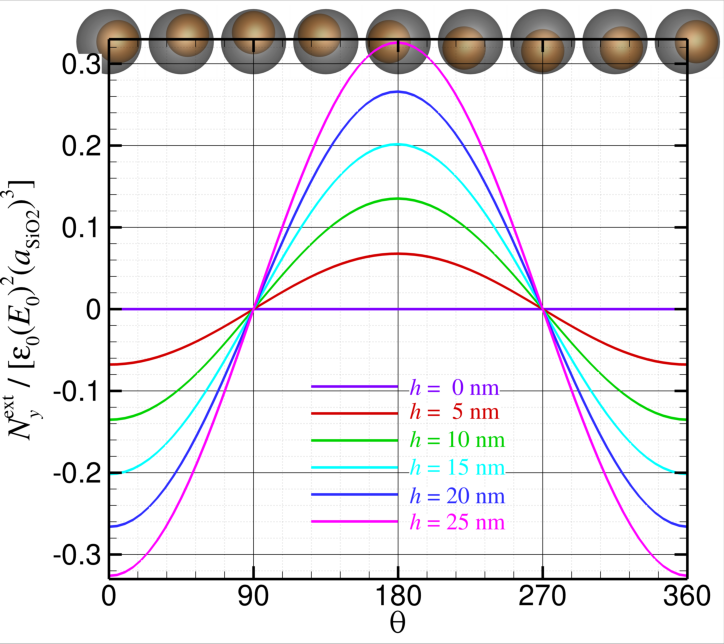}}
\caption{Optical torque perpendicular to the direction of incident beam propagation on the eccentric Au@SiO$_2$ core-shell particle under the linearly polarised Gaussian beam illumination. (a-c) in air with $n_{\text{air}}=1$; (d-f) in water with $n_{\text{water}}=1.33$. The shell radius is $a_{\text{shell}}=90$ nm, the core radius is $a_{\text{core}}=60$ nm, the beam wavelength is $\lambda=532$ nm, and the beam waist radius is $w_0=1$ $\mu$m.
}  \label{Fig:Ny_h}
\end{figure*}

\subsection{Optical torque}

\begin{figure*}[!ht]
\centering{}
\subfloat[Au@SiO$_{2}$]{ \includegraphics[width=0.32\textwidth]{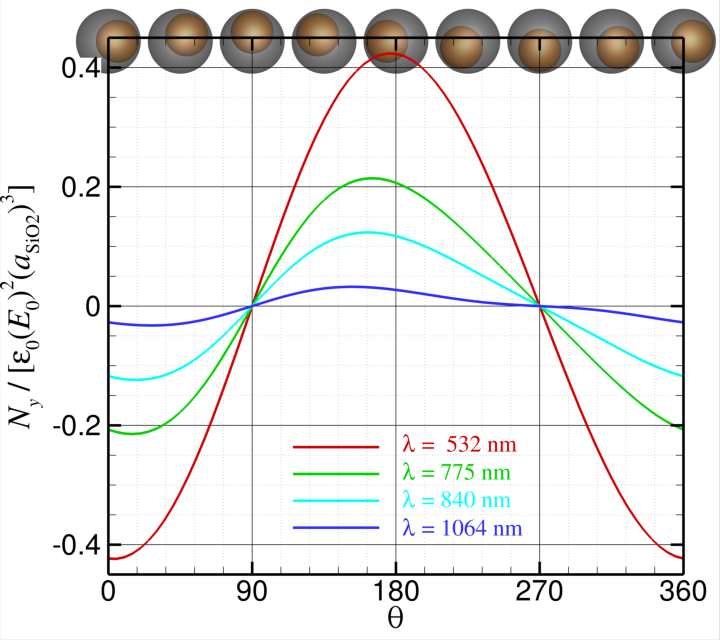}}
\subfloat[Au@SiO$_{2}$]{ \includegraphics[width=0.32\textwidth]{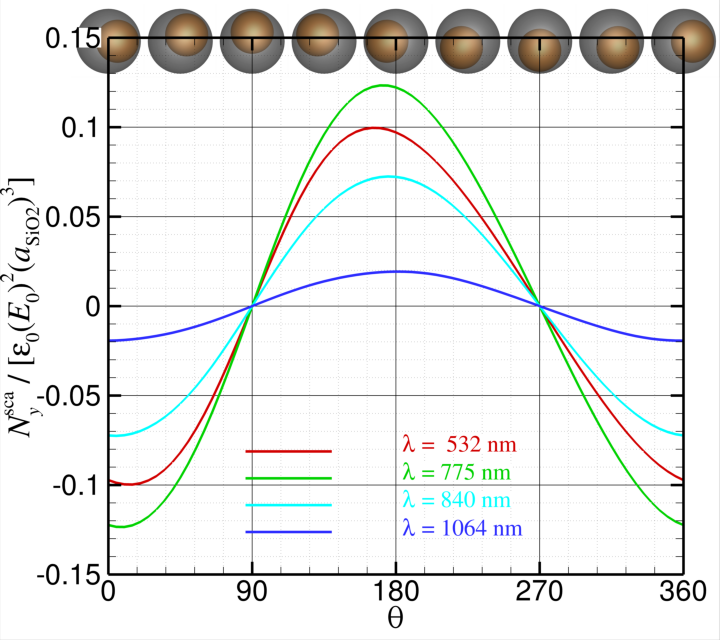}}
\subfloat[Au@SiO$_{2}$]{ \includegraphics[width=0.32\textwidth]{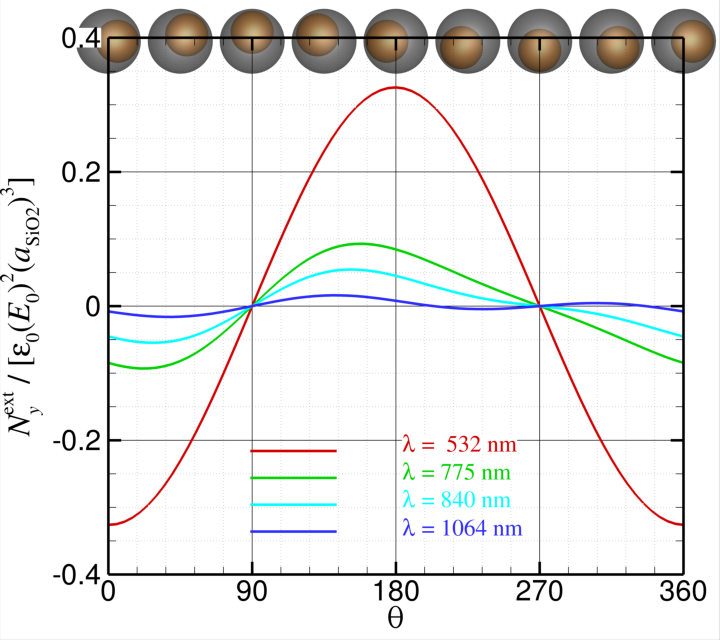}} \\
\subfloat[TiO$_2$@SiO$_{2}$]{ \includegraphics[width=0.32\textwidth]{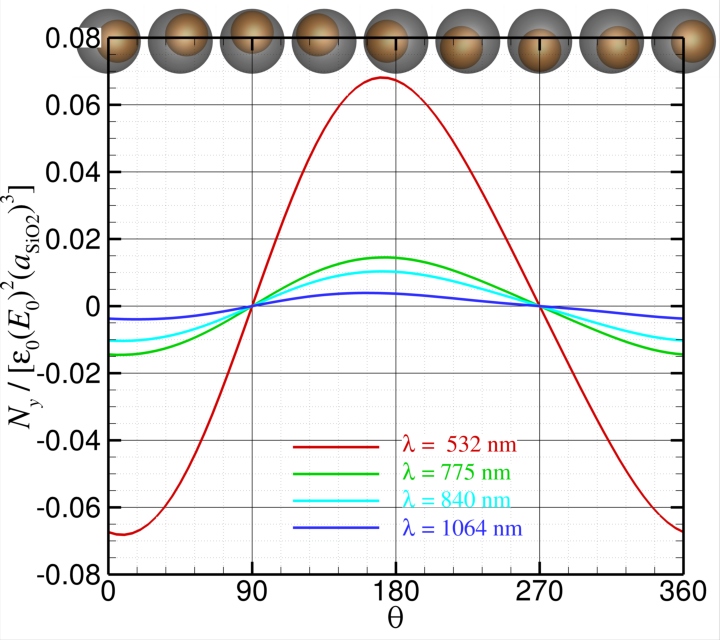}}
\subfloat[TiO$_2$@SiO$_{2}$]{ \includegraphics[width=0.32\textwidth]{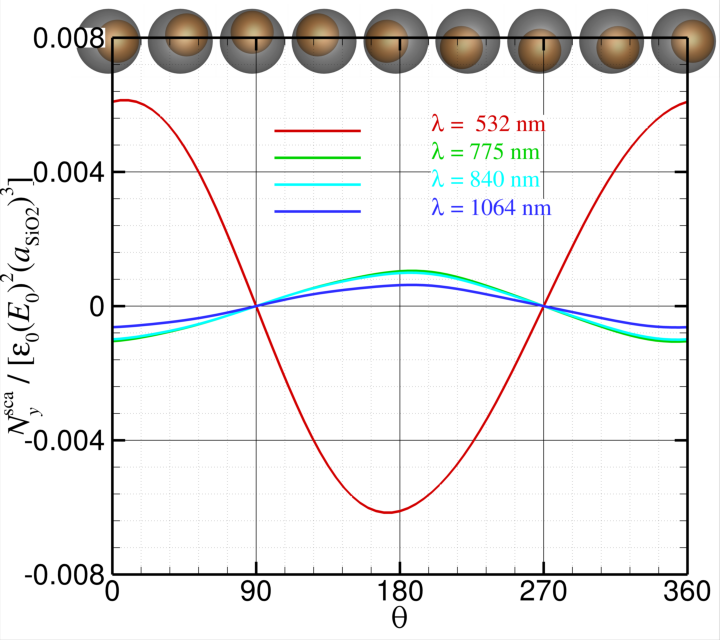}}
\subfloat[TiO$_2$@SiO$_{2}$]{ \includegraphics[width=0.32\textwidth]{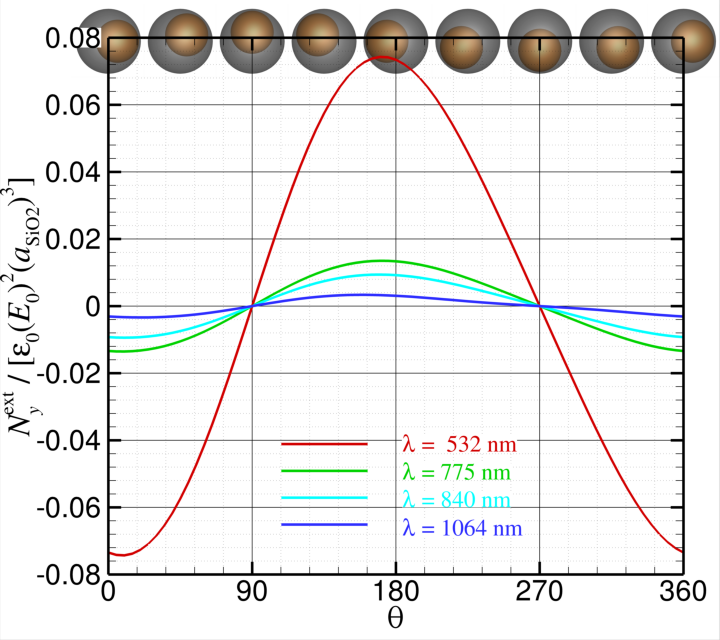}} \\
\subfloat[SiO$_2$@TiO$_{2}$]{ \includegraphics[width=0.32\textwidth]{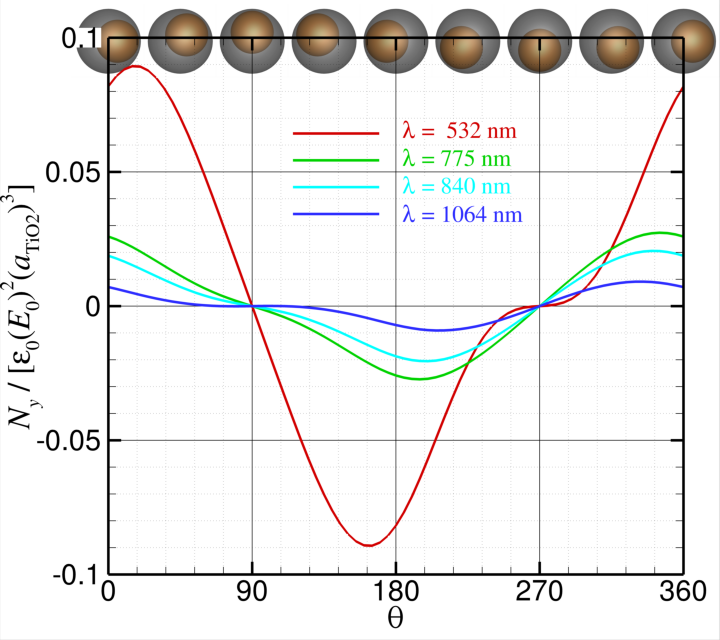}}
\subfloat[SiO$_2$@TiO$_{2}$]{ \includegraphics[width=0.32\textwidth]{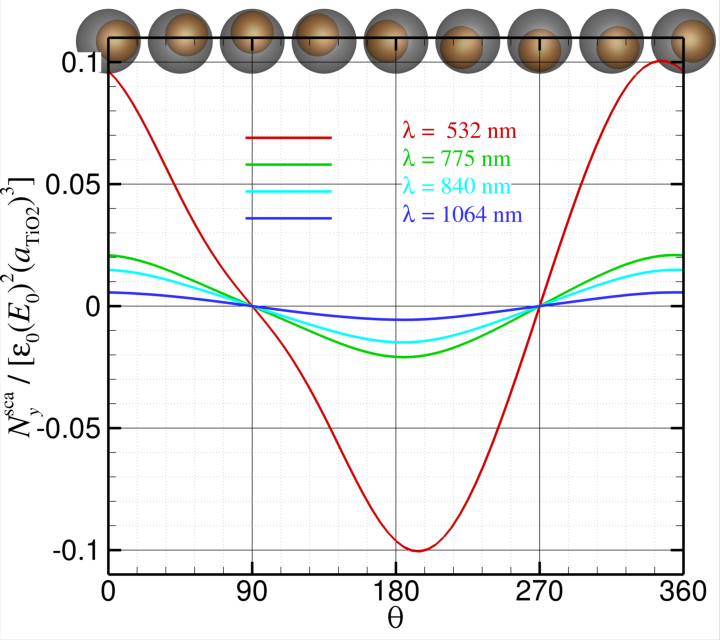}}
\subfloat[SiO$_2$@TiO$_{2}$]{ \includegraphics[width=0.32\textwidth]{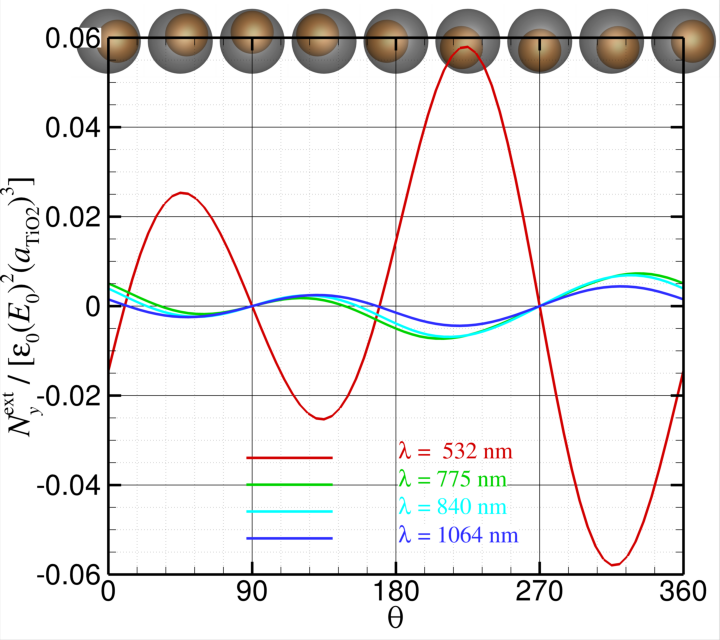}}
\caption{Optical torque perpendicular to the direction of incident beam propagation on three types of eccentric core-shell particles in water under a linearly polarised Gaussian beam illumination with beam waist radius as $w_0=1$ $\mu$m. The geometrical features of the core-shell particle are $a_{\text{shell}}=90$ nm, $a_{\text{core}}=60$ nm and $h=25$ nm.
}  \label{Fig:Ny_lambda}
\end{figure*}

The direction of the optical torque is perpendicular to the plane constructed by the light propagation direction and its electric field polarisation direction. Fig.~\ref{Fig:Ny_h} illustrates how the asymmetry, $h$ and orientation, $\theta$ of the Au core affect the optical torque on an eccentric Au@SiO$_2$ core-shell particle under the illumination of a linearly polarised Gaussian beam with wavelength as $\lambda=532$~nm. As shown in Figs.~\ref{Fig:Ny_h}(b-c) and (e-f), when the displacement $h$ between the centre of the Au core and that of the SiO$_2$ shell increases, the optical torques due to the scattered field and the interaction between the incident and scattered fields become more and more significant. However, these two effects are out of phase (180 degree difference) along the orientation angle of the Au core, $\theta$, when the surrounding medium is air, as shown in Figs.~\ref{Fig:Ny_h}(b) and (c). This leads to a small net optical torque as presented in Fig.~\ref{Fig:Ny_h}(a). When the surrounding medium is water, the optical torques due to the scattered field and the interaction between the incident and scattered fields are in phase along with the orientation angle of the Au core, $\theta$. As such, the total optical torque on the eccentric Au@SiO$_2$ core-shell particle in water is more significant relative to that in air.

Fig.~\ref{Fig:Ny_lambda} shows the effect of wavelength on the optical torque acting on three types of eccentric core-shell particles with $h=25$ nm under the illumination of a linearly polarised Gaussian beam in water. As displayed in the first row of Fig.~\ref{Fig:Ny_lambda}, for the Au@SiO$_2$ eccentric core-shell particle, the optical torque from the scattered field, $N_y^{\rmsc}$ and that from the interaction between the scattered and incident fields, $N_y^{\rmext}$ are in phase with respect to the orientation of the Au core, $\theta$, which leads to a significant net optical torque, $N_y$. As to the TiO$_2$@SiO$_{2}$ eccentric core-shell particle, net optical torque, $N_y$ is dominated by the interaction between the scattered and incident fields, $N_y^{\rmext}$; while for the SiO$_2$@TiO$_{2}$ eccentric core-shell particle, the contribution from the scattered field dominates. 

\begin{figure*}[t]
\centering{}
\subfloat[$F_x$]{ \includegraphics[width=0.3\textwidth]{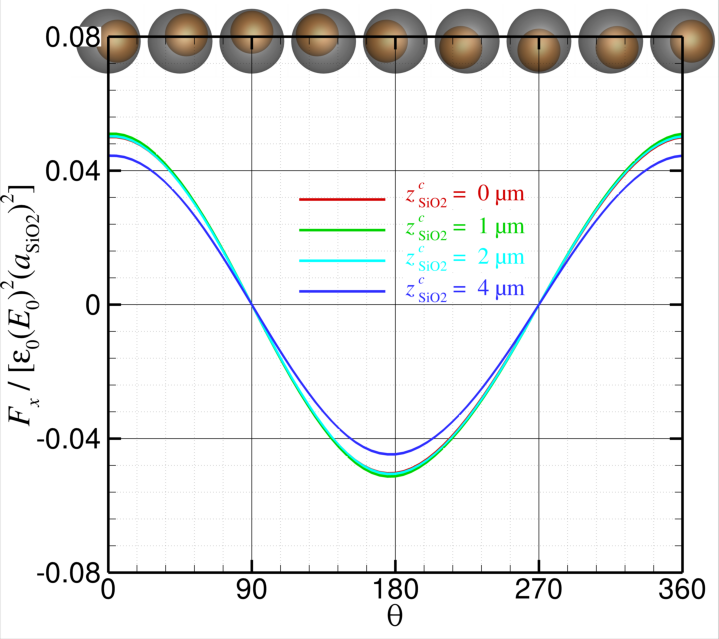}}
\subfloat[$F_z$]{ \includegraphics[width=0.3\textwidth]{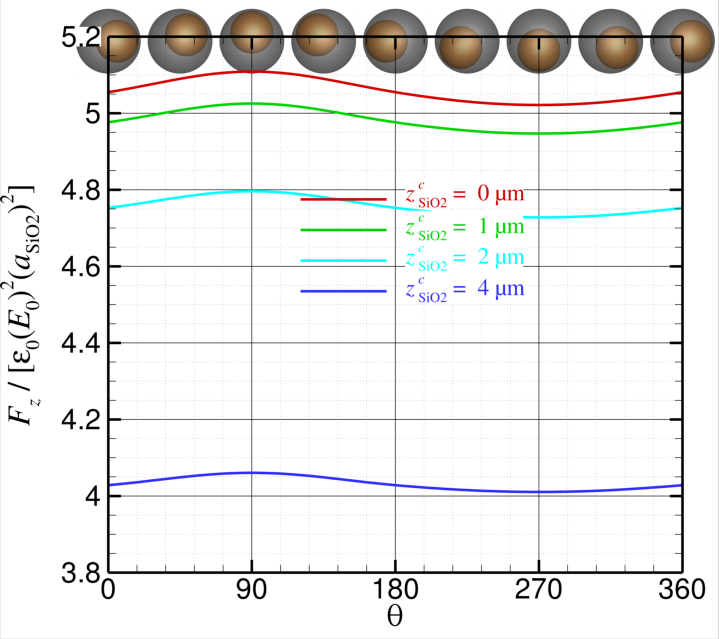}}
\subfloat[$N_y$]{ \includegraphics[width=0.3\textwidth]{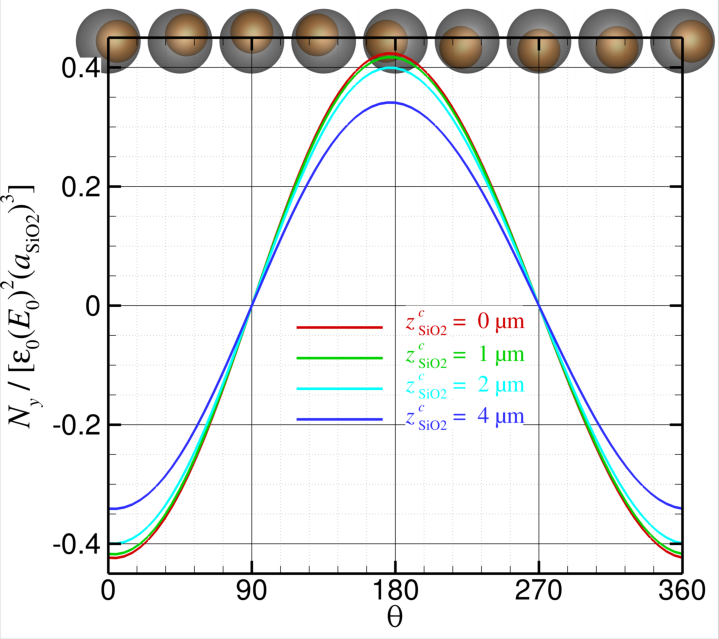}} \\
\subfloat[$F_x$]{ \includegraphics[width=0.3\textwidth]{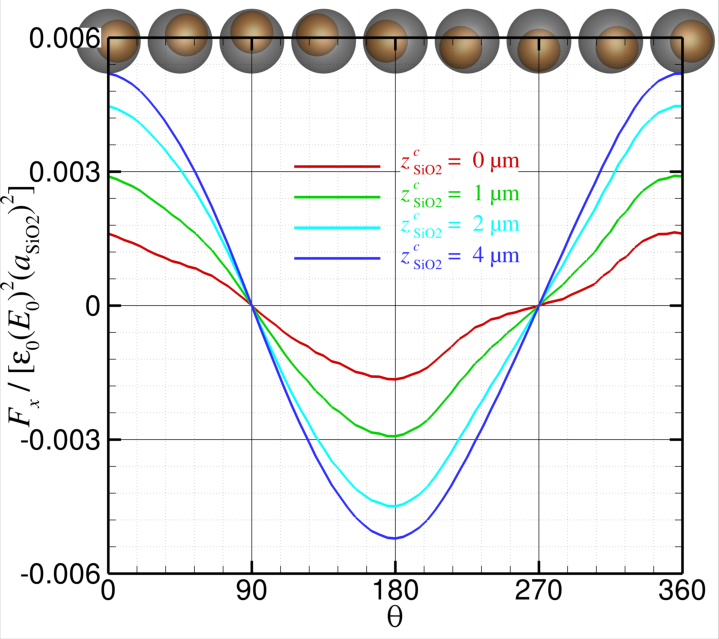}}
\subfloat[$F_z$]{ \includegraphics[width=0.3\textwidth]{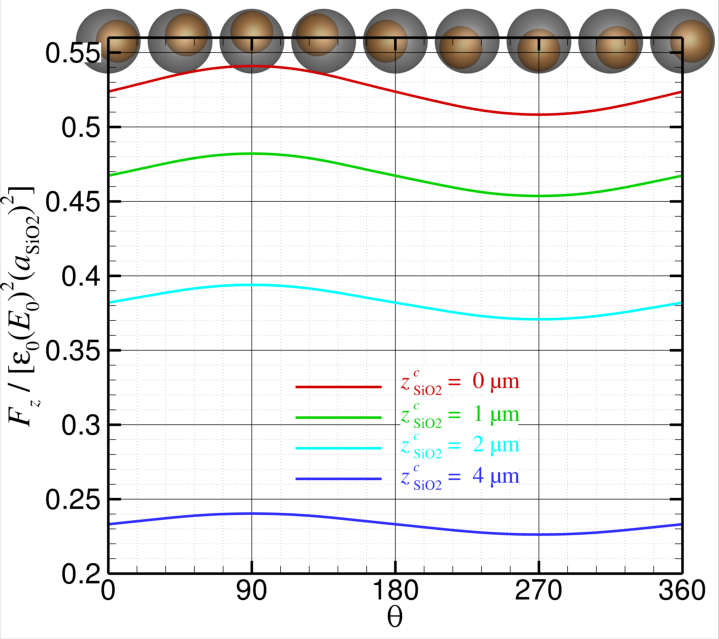}}
\subfloat[$N_y$]{ \includegraphics[width=0.3\textwidth]{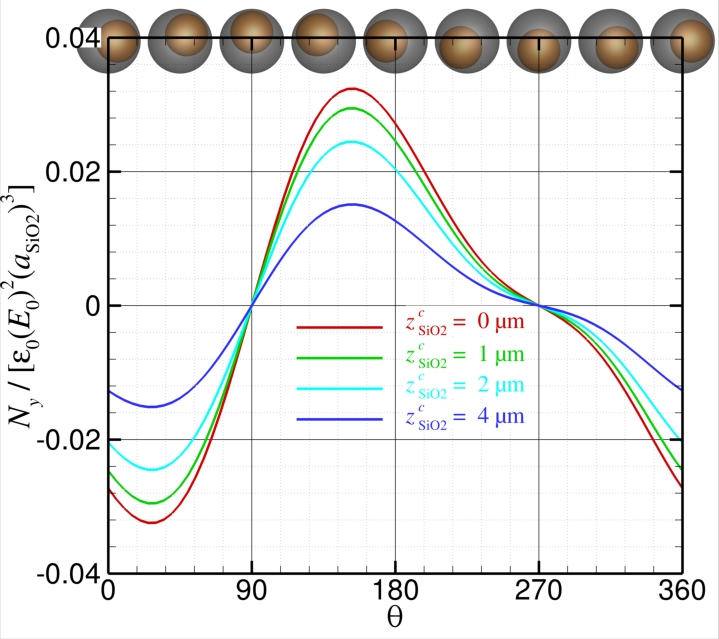}} 
\caption{Optical force and torque an eccentric Au@SiO$_2$ core-shell particle in water ($n_{\text{medium}} = 1.33$) under a circular polarised Gaussian beam illumination with beam waist radius as $w_0=1$ $\mu$m when the geometric centre of the shell locates at different positions along the beam propagation direction: (a-c) $\lambda$=532 nm and (d-f) $\lambda$=1064 nm. The geometric features of the eccentric core-shell particle are $a_{\text{shell}}=90$ nm, $a_{\text{core}}=60$ nm and $h=25$ nm.
}  \label{Fig:offcenAu@SiO2}
\end{figure*}

\subsection{When the eccentric core-shell particle is away from the focal point of the beam}

\begin{figure*}[t]
\centering{}
\subfloat[$F_x$]{ \includegraphics[width=0.3\textwidth]{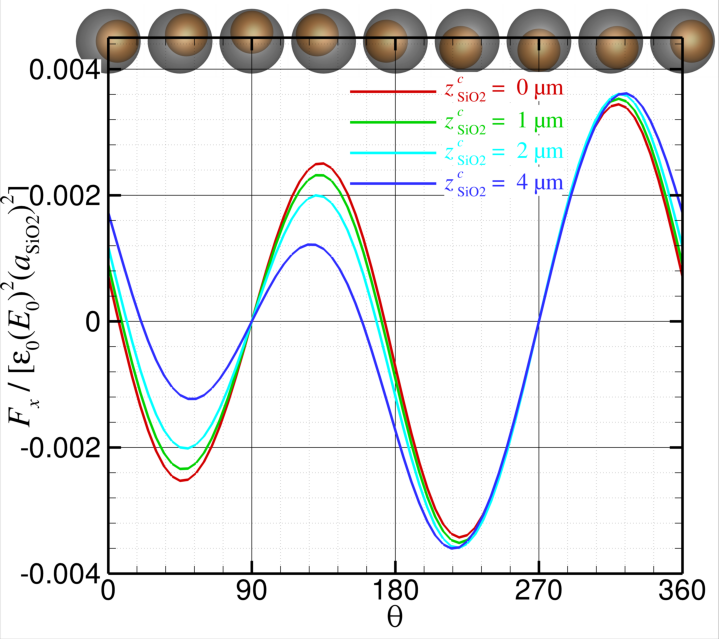}}
\subfloat[$F_z$]{ \includegraphics[width=0.3\textwidth]{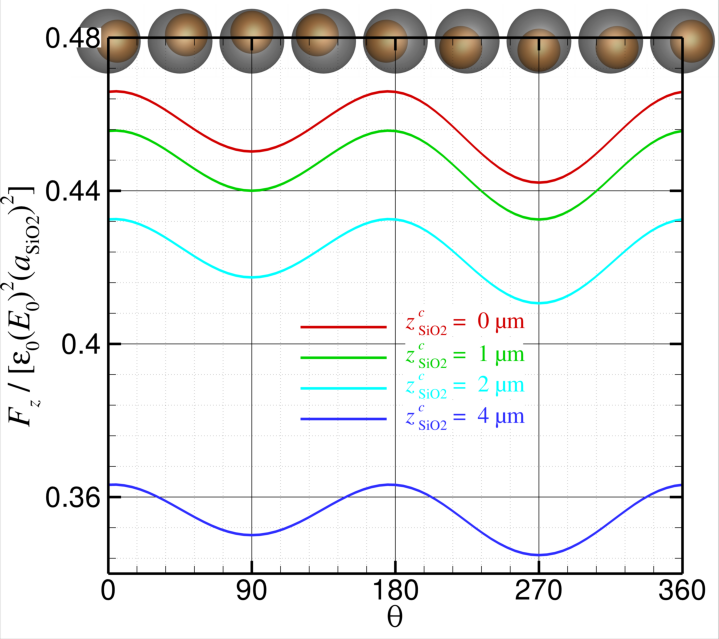}}
\subfloat[$N_y$]{ \includegraphics[width=0.3\textwidth]{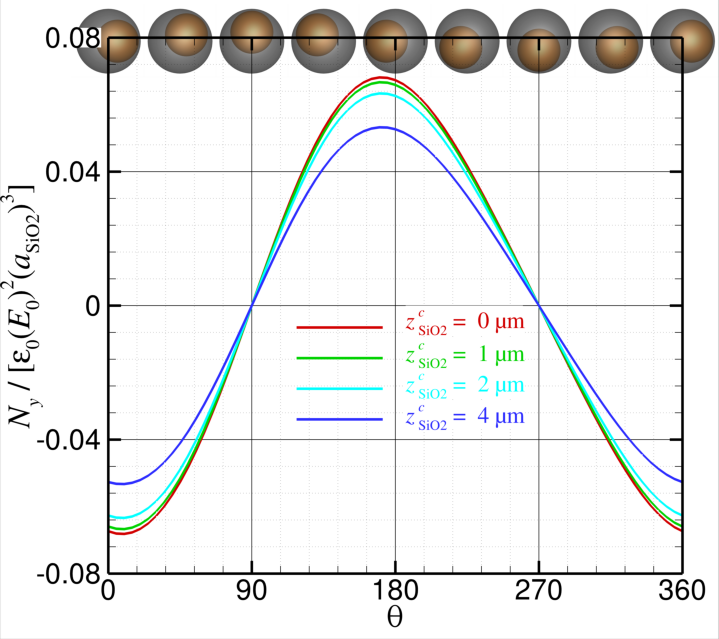}} \\
\subfloat[$F_x$]{ \includegraphics[width=0.3\textwidth]{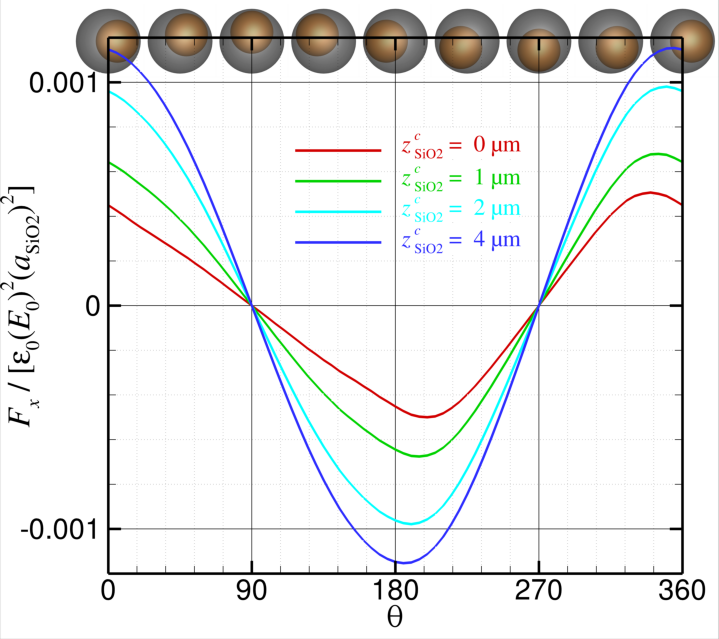}}
\subfloat[$F_z$]{ \includegraphics[width=0.3\textwidth]{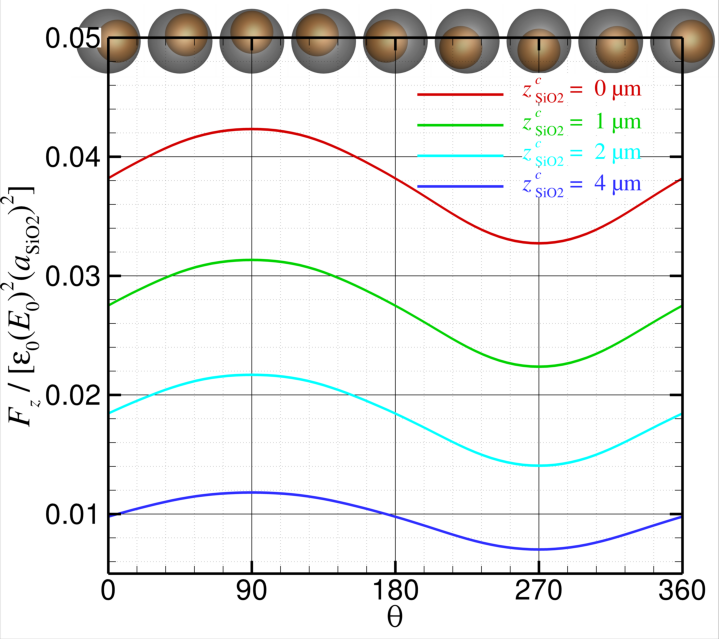}}
\subfloat[$N_y$]{ \includegraphics[width=0.3\textwidth]{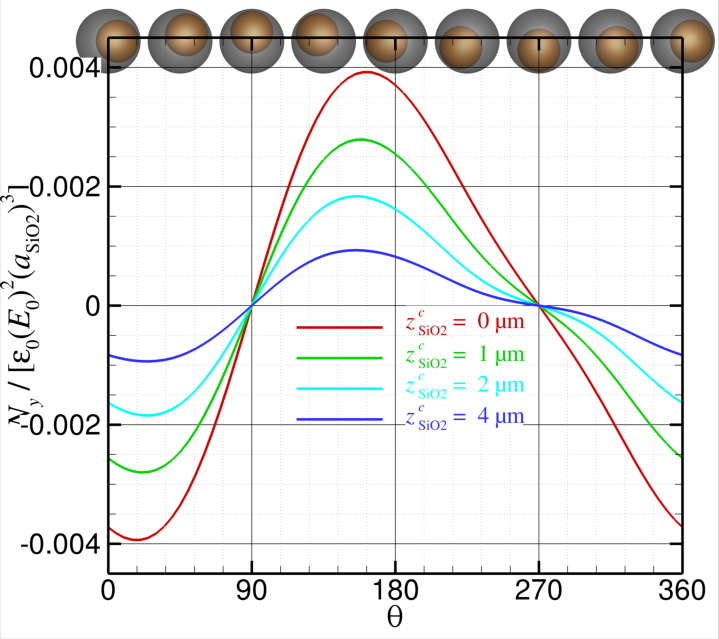}}
\caption{Optical force and torque an eccentric TiO$_2$@SiO$_2$ core-shell particle in water ($n_{\text{medium}} = 1.33$) under a circular polarised Gaussian beam illumination with beam waist radius as $w_0=1$ $\mu$m when the geometric centre of the shell locates at different positions along the beam propagation direction: (a-c) $\lambda$=532 nm and (d-f) $\lambda$=1064 nm. The geometric features of the eccentric core-shell particle are $a_{\text{shell}}=90$ nm, $a_{\text{core}}=60$ nm and $h=25$ nm.
}  \label{Fig:offcenTiO2@SiO2}
\end{figure*}

In the previous sections, we demonstrate the optomechanical response of an eccentric spherical core-shell particle under the linearly polarised Gaussian illumination when the centre of the shell is trapped at the focus of the beam. Nevertheless, in practice, the optical trapping location of the particle is not at the focal point of beam in most cases due to the field gradient. We then consider a few cases when the geometrical centre of the eccentric core-shell particle is at different positions relative to the beam focus along the beam propagation direction, denoted as $(0,0,z^c_{\text{shell}})$. From Figs.~\ref{Fig:offcenAu@SiO2} and \ref{Fig:offcenTiO2@SiO2}, we can see that the magnitude of the trapping force $F_z$ changes significantly when $z^c_{\text{shell}}$ varies from $z^c_{\text{shell}}=0$ $\mu$m to $z^c_{\text{shell}}=1$ $\mu$m, $z^c_{\text{shell}}=2$ $\mu$m and $z^c_{\text{shell}}=4$ $\mu$m, while the magnitude of the optical torque $N_y$ does not show an obvious variation. This indicates that the optical rotation of nanoscale eccentric core-shall particles should be observable in realistic experiments using current technology.

\section{Optical torques on an eccentric spherical core-shell particle under a Gaussian beam with circular polarisation}

Circularly polarised Gaussian beams have commonly been used to optically rotate microscale birefringent particles. In this section, we explore the optical torques acting on an eccentric core-shell particle under the Gaussian illumination with circular polarisation. Suppose a circularly polarised beam with Gaussian profile propagates along $z$ direction, such a beam can be described as:
\begin{subequations}\label{eq:EHinc_circular}
    \begin{align}
        E_{x}^{\rminc} = & \quad \frac{\sqrt{2}}{2} E_{0} \bigg \{ 1 + s^2(-\varrho^2 \vartheta^2 - \rmi \varrho^4 \vartheta^3 - 2\vartheta^2 \xi^2)  \nonumber \\
        & \qquad \quad +  s^4 \left[ 2\varrho^4\vartheta^4 + 3\rmi \varrho^6 \vartheta^5 - 0.5 \varrho^8 \vartheta^6 + (8 \varrho^2\vartheta^4 + 2 \rmi \varrho^4 \vartheta^5) \xi^2 \right] \bigg \}\psi_0 e^{\rmi kz} \nonumber \\
        &  + \frac{\rmi \sqrt{2}}{2} E_{0} \left\{s^2(-2\vartheta^2 \xi \eta) + s^4 \left[ 8\varrho^2\vartheta^4 + 2\rmi \varrho^4 \vartheta^5) \xi \eta \right]
        \right\}\psi_0 e^{\rmi kz},
    \end{align}
    \begin{align}
        E_{y}^{\rminc} = & \quad \frac{\sqrt{2}}{2} E_{0} \left\{s^2(-2\vartheta^2 \xi \eta) + s^4 \left[ 8\varrho^2\vartheta^4 + 2\rmi \varrho^4 \vartheta^5) \xi \eta \right]
        \right\}\psi_0 e^{\rmi kz} \nonumber \\
        &  + \frac{\rmi \sqrt{2}}{2} E_{0}  \bigg \{ 1 + s^2(-\varrho^2 \vartheta^2 - \rmi \varrho^4 \vartheta^3 - 2\vartheta^2 \eta^2)  \nonumber \\
        & \qquad \quad +  s^4 \left[ 2\varrho^4\vartheta^4 + 3\rmi \varrho^6 \vartheta^5 - 0.5 \varrho^8 \vartheta^6 + (8 \varrho^2\vartheta^4 + 2 \rmi \varrho^4 \vartheta^5) \eta^2 \right] \bigg \}\psi_0 e^{\rmi kz},
    \end{align}
    \begin{align}
        E_{z}^{\rminc} = & \quad \frac{\sqrt{2}}{2} E_{0} \left\{s(-2\vartheta \xi) + s^3\left[ (6\varrho^2 \vartheta^3 +2 \rmi \varrho^4 \vartheta^4)\xi \right] \right. \nonumber \\
        & \qquad + \left. s^5 \left[ -20\varrho^4\vartheta^5 - 10\rmi \varrho^6 \vartheta^6 + \varrho^8 \vartheta^7   \right] \xi
        \right\}\psi_0 e^{\rmi kz} \nonumber \\
        & + \frac{\rmi \sqrt{2}}{2} E_{0}  \left\{s(-2\vartheta \eta) + s^3\left[ (6\varrho^2 \vartheta^3 +2 \rmi \varrho^4 \vartheta^4)\eta \right] \right. \nonumber \\
        & \qquad + \left. s^5 \left[ -20\varrho^4\vartheta^5 - 10\rmi \varrho^6 \vartheta^6 + \varrho^8 \vartheta^7   \right] \eta
        \right\}\psi_0 e^{\rmi kz}, 
    \end{align}
    \begin{align}
        H_{x}^{\rminc} = & \quad \frac{\sqrt{2}}{2} \frac{k}{\mu_0 \mu \omega} E_{0} \left\{s^2(-2\vartheta^2 \xi \eta) + s^4 \left[ 8\varrho^2\vartheta^4 + 2\rmi \varrho^4 \vartheta^5) \xi \eta \right]
        \right\}\psi_0 e^{\rmi kz} \nonumber \\
        &  - \frac{\rmi \sqrt{2}}{2} \frac{k}{\mu_0 \mu \omega} E_{0}  \bigg \{ 1 + s^2(-\varrho^2 \vartheta^2 - \rmi \varrho^4 \vartheta^3 - 2\vartheta^2 \xi^2)  \nonumber \\
        & \qquad \quad +  s^4 \left[ 2\varrho^4\vartheta^4 + 3\rmi \varrho^6 \vartheta^5 - 0.5 \varrho^8 \vartheta^6 + (8 \varrho^2\vartheta^4 + 2 \rmi \varrho^4 \vartheta^5) \xi^2 \right] \bigg \}\psi_0 e^{\rmi kz},
    \end{align}
    \begin{align}
        H_{y}^{\rminc} = & \quad \frac{\sqrt{2}}{2} \frac{k}{\mu_0 \mu \omega} E_{0} \bigg \{ 1 + s^2(-\varrho^2 \vartheta^2 - \rmi \varrho^4 \vartheta^3 - 2\vartheta^2 \eta^2)  \nonumber \\
        & \qquad \quad +  s^4 \left[ 2\varrho^4\vartheta^4 + 3\rmi \varrho^6 \vartheta^5 - 0.5 \varrho^8 \vartheta^6 + (8 \varrho^2\vartheta^4 + 2 \rmi \varrho^4 \vartheta^5) \eta^2 \right] \bigg \}\psi_0 e^{\rmi kz} \nonumber \\
        &  - \frac{\rmi \sqrt{2}}{2} \frac{k}{\mu_0 \mu \omega} E_{0} \left\{s^2(-2\vartheta^2 \xi \eta) + s^4 \left[ 8\varrho^2\vartheta^4 + 2\rmi \varrho^4 \vartheta^5) \xi \eta \right]
        \right\}\psi_0 e^{\rmi kz},
    \end{align}
    \begin{align}
        H_{z}^{\rminc} = & \quad \frac{\sqrt{2}}{2} \frac{k}{\mu_0 \mu \omega} E_{0} \left\{s(-2\vartheta \eta) + s^3\left[ (6\varrho^2 \vartheta^3 +2 \rmi \varrho^4 \vartheta^4)\eta \right] \right. \nonumber \\
        & \qquad + \left. s^5 \left[ -20\varrho^4\vartheta^5 - 10\rmi \varrho^6 \vartheta^6 + \varrho^8 \vartheta^7   \right] \eta
        \right\}\psi_0 e^{\rmi kz} \nonumber \\
        & - \frac{\rmi \sqrt{2}}{2} \frac{k}{\mu_0 \mu \omega} E_{0}  \left\{s(-2\vartheta \xi) + s^3\left[ (6\varrho^2 \vartheta^3 +2 \rmi \varrho^4 \vartheta^4)\xi \right] \right. \nonumber \\
        & \qquad + \left. s^5 \left[ -20\varrho^4\vartheta^5 - 10\rmi \varrho^6 \vartheta^6 + \varrho^8 \vartheta^7   \right] \xi
        \right\}\psi_0 e^{\rmi kz}, 
    \end{align}
\end{subequations}
Using Eq.~(\ref{eq:EHinc_circular}) as the incident field and following the simulation demonstrated in Sec. 2 of the main text, we can get the optomechanical response of an eccentric core-shell particle under the illumination of a Gaussian beam with circular polarisation. 

\begin{figure*}[!ht]
\centering{}
\subfloat[$a_{\text{shell}}=90$ nm, $a_{\text{core}}=60$ nm, $h=25$ nm]{ \includegraphics[width=0.32\textwidth]{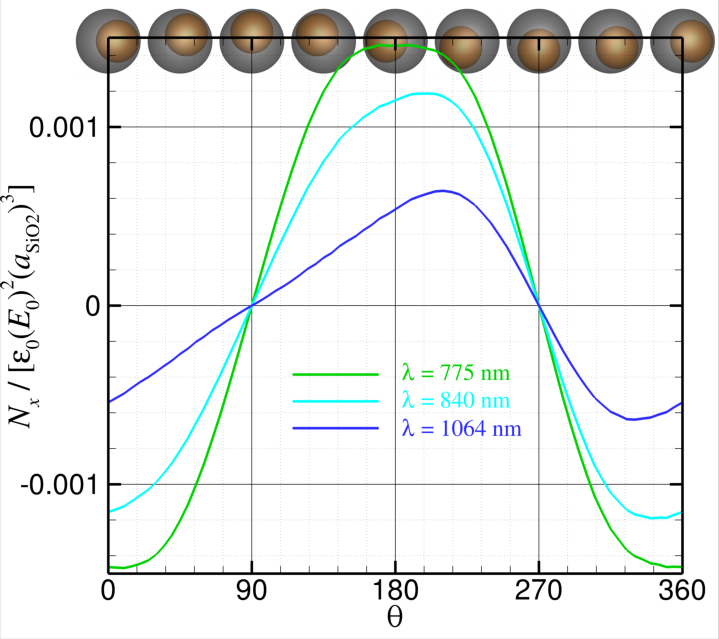}}
\subfloat[$a_{\text{shell}}=90$ nm, $a_{\text{core}}=60$ nm, $h=10$ nm]{ \includegraphics[width=0.32\textwidth]{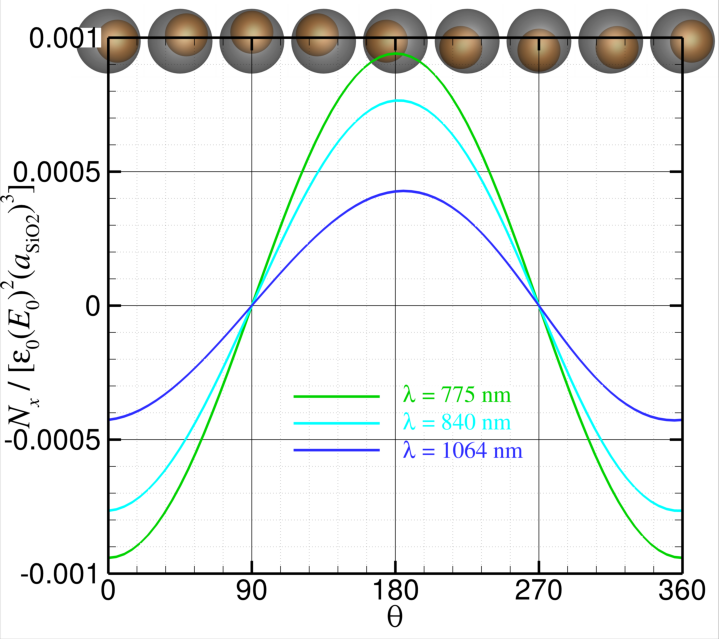}}
\subfloat[$a_{\text{shell}}=45$ nm, $a_{\text{core}}=30$ nm, $h=10$ nm]{ \includegraphics[width=0.32\textwidth]{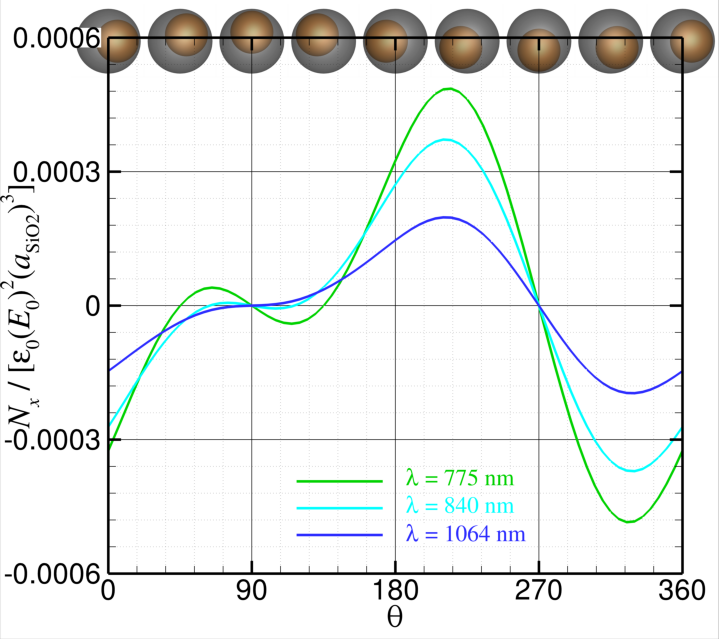}} \\
\subfloat[$a_{\text{shell}}=90$ nm, $a_{\text{core}}=60$ nm, $h=25$ nm]{ \includegraphics[width=0.32\textwidth]{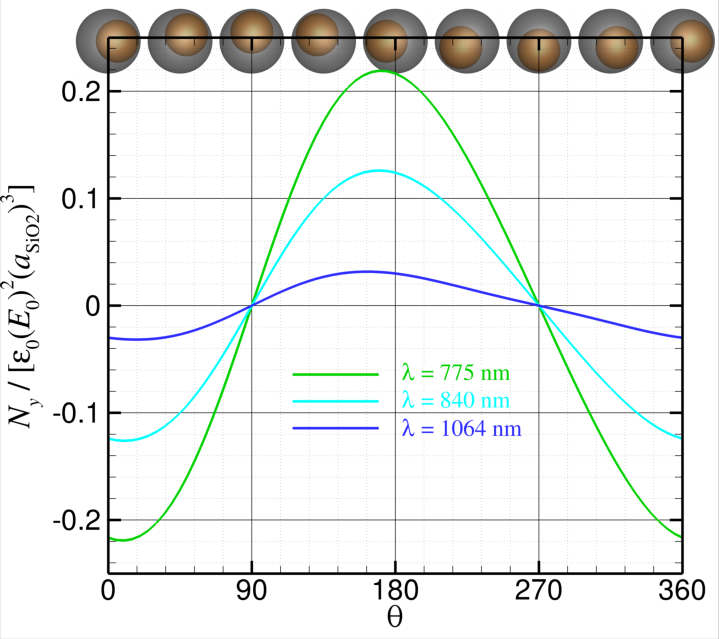}}
\subfloat[$a_{\text{shell}}=90$ nm, $a_{\text{core}}=60$ nm, $h=10$ nm]{ \includegraphics[width=0.32\textwidth]{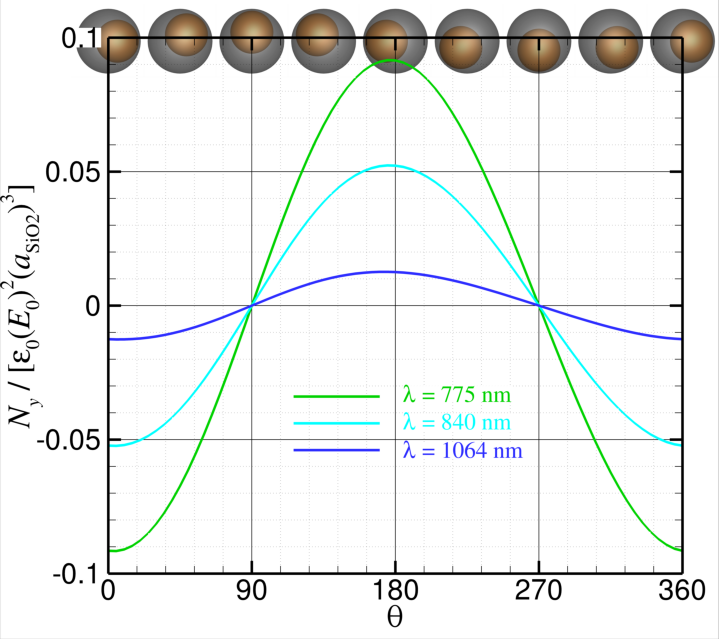}}
\subfloat[$a_{\text{shell}}=45$ nm, $a_{\text{core}}=30$ nm, $h=10$ nm]{ \includegraphics[width=0.32\textwidth]{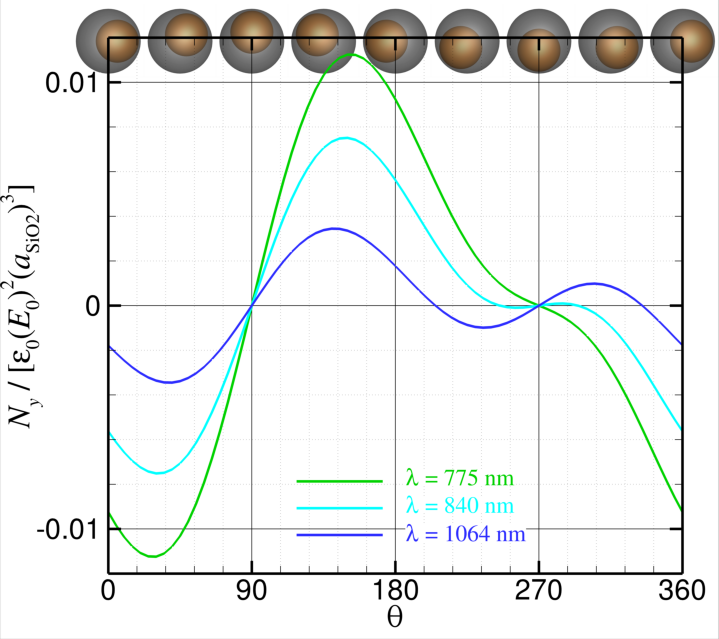}} \\
\subfloat[$a_{\text{shell}}=90$ nm, $a_{\text{core}}=60$ nm, $h=25$ nm]{ \includegraphics[width=0.32\textwidth]{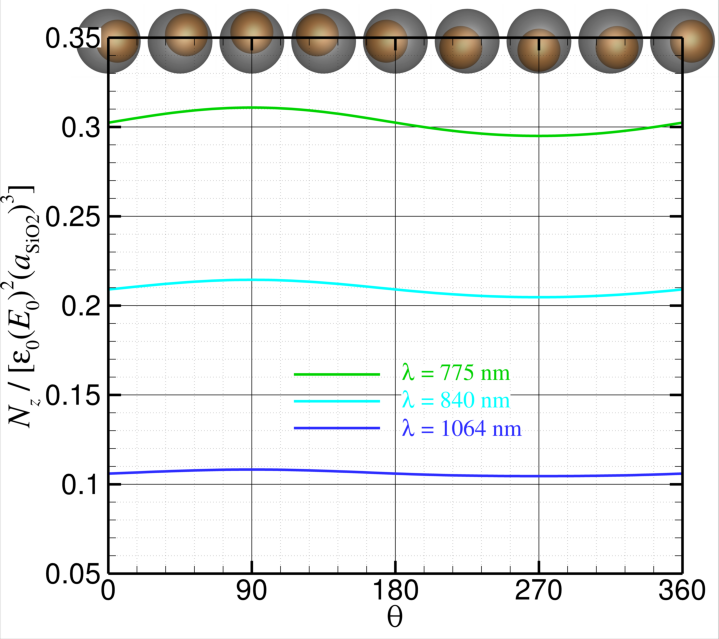}}
\subfloat[$a_{\text{shell}}=90$ nm, $a_{\text{core}}=60$ nm, $h=10$ nm]{ \includegraphics[width=0.32\textwidth]{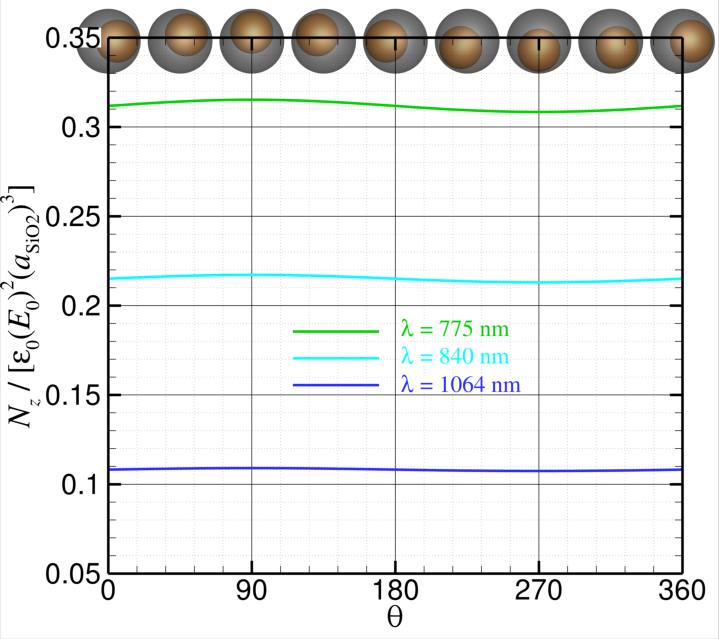}}
\subfloat[$a_{\text{shell}}=45$ nm, $a_{\text{core}}=30$ nm, $h=10$ nm]{ \includegraphics[width=0.32\textwidth]{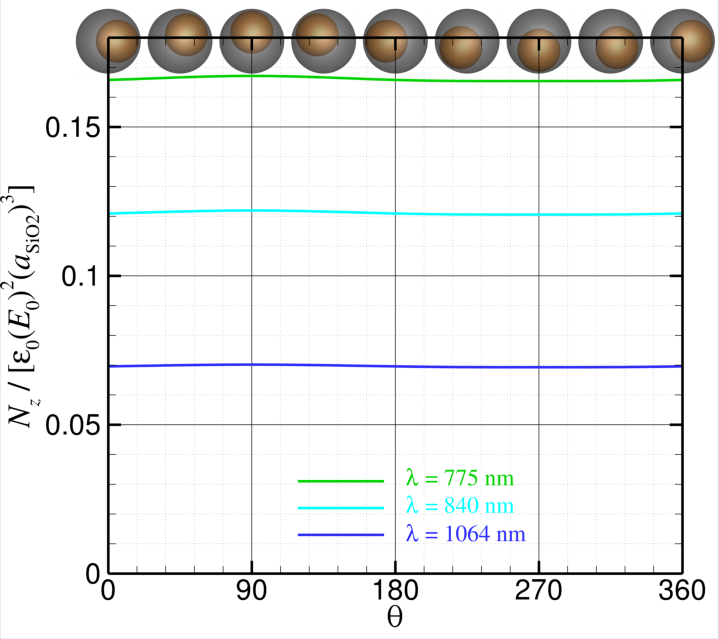}}
\caption{Optical torques on an eccentric Au@SiO$_2$ core-shell particle in water with $n_{\text{medium}} = 1.33$ under a circular polarised Gaussian beam illumination with beam waist radius as $w_0=1$ $\mu$m.
}  \label{Fig:Ncp_AuSiO2}
\end{figure*}

As discussed in the main text, to compare with the case of birefringent particles, we calculated the optical torque on a nanoscale eccentric spherical core-shell particle which is composed by a dielectric core and a dielectric shell, such as TiO$_2$@SiO$_2$ and SiO$_2$@TiO$_2$. Our results show that the optical torque that is perpendicular the beam direction due to the asymmetry of the eccentric spherical core-shell particle is much higher than the optical torque along the wave propagation direction due to the circular polarisation of the beam. 

We also consider another type of eccentric spherical core-shell particle that has a metallic core in a dielectric shell, for example Au@SiO$_2$. Fig.~\ref{Fig:Ncp_AuSiO2} demonstrates how the optical torques acting on an eccentric spherical Au@SiO$_2$ core-shell particle in water ($n_{\text{medium}}=1.33$) change with respect to the orientation of the Au core, $\theta$, when the geometrical centre of the SiO$_2$ shell is trapped at the beam focus. Due to the eccentricity and refractive index profile, torques $N_x$ and $N_y$ appear and vary with respect to $\theta$. In our simulations, as the eccentric feature (asymmetry), $\theta$, varies in the $xz$ plane, $N_y$ is the dominating torque relative to $N_x$, as shown in the first and second rows of Fig.~\ref{Fig:Ncp_AuSiO2}. Comparing Fig.~\ref{Fig:Ncp_AuSiO2} (d) and Fig.~\ref{Fig:Ny_lambda} (a), we can see that the amplitude of $N_y$ under the circularly polarised Gaussian illumination is in the same order as that under the linearly polarised Gaussian beam. Due to the circular polarisation, the torque along the wave propagation, $N_z$, also appears. As shown in the third row of Fig.~\ref{Fig:Ncp_AuSiO2}, the magnitude of $N_z$ is in the same order as $N_y$. Such an enhanced torque $N_z$ acting on the Au@SiO$_2$ eccentric spherical core-shell particle relative to that on the TiO$_2$@SiO$_2$ eccentric spherical core-shell particle is mainly due to the strong absorption of the Au core at the wavelengths under consideration. We would like also to mention that the noticeable optical torque $N_z$ acting on such a Au@SiO$_2$ particle by using the circular polarisation beam can be achievable when using the concentric core-shell particle, as listed in Table~\ref{tab:Au@SiO2con}. Nevertheless, the rotation of a fully symmetric particle is hard to observe from its steady scattered pattern at the far field, which makes it not an ideal probe for the local measurement of nanoscale environments. 

\begin{table*}[t]
\centering
\caption{\bf Non-dimensional optical torque $\hat{N}_z \equiv N_z/[\epsilon_0 E_0^2 (a_{\text{SiO}_{2}})^3]$ on concentric Au@SiO$_2$ core-shell particle ($h=0$) under the illumination of Gaussian beam with circular polarisation and waist radius as 1$\mu$m in water at different wavelengths, $\lambda$, and the corresponding rotation frequency $\Omega$ (in Hz) when the beam power is $P_0 = $ 20 mW.}
\resizebox{1.0\textwidth}{!}{
\begin{tabular}{l | c | c | c | c | c | c | c | c | c | c | c | c | c | c | c | c }
\hline
& \multicolumn{8}{c|} {$a_\text{shell}$=90 nm, $a_\text{shell}$=45 nm} & \multicolumn{8}{c} {$a_\text{shell}$=60 nm, $a_\text{shell}$=30 nm} \\
\cline{2-17}
& \multicolumn{2}{c|} {$z^{c}$=0 $\mu$m} & \multicolumn{2}{c|} {$z^{c}$=1 $\mu$m} & \multicolumn{2}{c|} {$z^{c}$=2 $\mu$m} & \multicolumn{2}{c|} {$z^{c}$=4 $\mu$m} & \multicolumn{2}{c|} {$z^{c}$=0 $\mu$m} & \multicolumn{2}{c|} {$z^{c}$=1 $\mu$m} & \multicolumn{2}{c|} {$z^{c}$=2 $\mu$m} & \multicolumn{2}{c} {$z^{c}$=4 $\mu$m} \\
\cline{2-17}
& $\hat{N}_z$ & $\Omega$ & $\hat{N}_z$ & $\Omega$ & $\hat{N}_z$ & $\Omega$ & $\hat{N}_z$ & $\Omega$ & $\hat{N}_z$ & $\Omega$ & $\hat{N}_z$ & $\Omega$ & $\hat{N}_z$ & $\Omega$ & $\hat{N}_z$ & $\Omega$ \\
\hline
$\lambda$=775 nm  & 0.313 & 594 & 0.303 & 575 & 0.276 & 523 & 0.202 & 383 & 0.169 & 314 & 0.163 & 303  & 0.148 & 275  & 0.109 & 202 \\
$\lambda$=840 nm  & 0.216 & 407 & 0.208 & 392 & 0.186 & 350 & 0.131 & 247 & 0.123 & 227 & 0.118 & 218  & 0.106 & 195  & 0.075 & 138 \\
$\lambda$=1064 nm & 0.109 & 205 & 0.102 & 192 & 0.086 & 162 & 0.053 & 99  & 0.070 & 125 & 0.066 & 118 & 0.056 & 100 & 0.035 & 62 \\
\hline
\end{tabular}
}
\label{tab:Au@SiO2con}
\end{table*}